\documentclass[twocolumn,showpacs,preprintnumbers,amsmath,amssymb,superscriptaddress]{revtex4}
\usepackage{longtable}
\usepackage{graphicx}
\usepackage{dcolumn}
\usepackage{bm}

\begin{document}

\title {Measuring Muon-Induced Neutrons with Liquid Scintillation Detector 
		at Soudan Mine}
\newcommand{\usd}{Department of Physics, The University of South Dakota, Vermillion, South Dakota 57069}
\newcommand{\tgu}{College of Sciences, China Three Gorges University, Yichang 443002, China}
\newcommand{\yru}{School of Physics and Optoelectronic Engineering, Yangtze University, Jingzhou 434023, China}
\newcommand{\dmm}{Dongming.Mei@usd.edu}

\author{C. Zhang} \affiliation{ \usd} \affiliation{\tgu}
\author{D.-M. Mei} \altaffiliation[Corresponding Author: ]{\dmm}  \affiliation{   \usd    } \affiliation{   \yru   }

\begin{abstract}
We report a direct detection of muon-induced high energy neutrons 
with a 12-liter neutron detector 
fabricated with EJ-301 liquid scintillator operating at Soudan Mine for about two years. The detector response
to energy from a few MeV up to $\sim$ 20 MeV has been calibrated using radioactive sources and cosmic-ray muons. 
Subsequently, we have calculated the scintillation efficiency for nuclear recoils, up to a few hundred MeV, 
using Birks' law in the Monte Carlo simulation. 
Data from an exposure of 655.1 days were analyzed and neutron-induced recoil events were observed in the 
energy region from 4 MeV to
50 MeV, corresponding to fast neutrons with kinetic energy
up to a few hundred MeV, depending on the scattering angle. Combining with the Monte Carlo simulation,
the measured muon-induced fast neutron flux is determined to be 
 $(2.23 \pm 0.52 (sta.) \pm 0.99 (sys.) ) \times10^{-9}$ cm$^{-2}$s$^{-1}$ (E$_{n}$ $>20$ MeV),
in a reasonable agreement with the model prediction. 
The muon flux is found to be
($1.65\pm 0.02 (sta.) \pm 0.1 (sys.) ) \times10^{-7}$ cm$^{-2}$s$^{-1}$ (E$_{\mu}$ $>$ 1 GeV),
consistent with other measurements. As a result, the muon-induced high energy gamma-ray flux 
is simulated to be 7.08 $\times$10$^{-7}$cm$^{-2}$s$^{-1}$ (E$_{\gamma}$ $>$ 1 MeV) 
for the depth of Soudan. 
\end{abstract}

\maketitle
\section{Introduction}
Measuring muon-induced fast neutrons is important to the understanding of backgrounds for
many rare event physics experiments including direct searches for dark matter.
Dark matter is believed to account for about a quarter of the mass-energy 
budget of the known universe~\cite{trimble}. However, the nature of dark 
matter is still mysterious to us so far. As a candidate of 
dark matter, Weakly Interacting Massive Particle (WIMP) is 
a target for direct detection through a set of underground experiments.   
Some of them, such as DAMA~\cite{dama}, 
CDMS-Si~\cite{cdms}, CoGeNT~\cite{cogent}, and CRESST-II~\cite{cresst} have claimed positive results while
others including CDMS-Ge, Xenon100, LUX~\cite{cdms2, xenon100, lux}, and SuperCDMS~\cite{supercdms}
 have ruled out those claims. Many experiments have 
set an upper limit on the mass cross-section contour
of WIMPs interacting with normal 
matters~\cite{picasso, naiad, zeplin, edelweiss, simple, superKamiokande, cdms2, xenon100, lux, supercdms}. The next
generation ton-scale experiments aim to achieve a sensitivity of $\sim$10$^{-48}$ cm$^{2}$ to WIMP-Nucleon cross-section
for WIMP mass of $\sim$100 GeV.  

Understanding the background events is the key to the success of any dark matter search experiment. 
Because they behave in a manner similar  to WIMPs, fast neutrons are taken as a vital background for 
these rare event physics experiments at deep underground.
Although the cosmogenic effects are dramatically suppressed 
by rock overburden~\cite{mei}, 
the energy spectrum, angular and 
multiplicity distribution of the fast neutrons induced by muons underground are not well understood~\cite{yfw, vku, mei, hav, vkl, ggv,mma, har}.
Muon-induced neutron production rates in different targets have been recently measured by many experiments~\cite{sbe, borx, lre} 
through measuring neutron captures. The direct measurements of neutron energy spectrum have not yet been reported. 
The muon-induced fast neutrons with energy above $\sim$10 MeV are difficult to shield and can contribute to the total background budget 
for a given experiment. The fast neutrons, from $(\alpha, n)$ reaction and fission decay in the surrounding rocks, are lower in energy
and thus easier to stop. 
To characterize those neutrons as a source of background for dark matter experiments in an underground environment 
by deploying a neutron detector in-situ will definitely help the understanding of the experimental results.        
\par
A neutron measurement usually involves identity discrimination using
scintillation detectors and energy scaling utilizing the time of flight (TOF)
technique. However, the TOF measurement will largely limit the 
acceptance of neutrons. Because of the low neutron intensity at deep underground sites, 
a neutron detector with large detection efficiency is needed.  
In addition to the TOF technique, the recoil energy of ions in liquid scintillators
can represent the energy of incident neutrons if the detector response to nuclear recoils is well 
understood with a Monte Carlo simulation. 
The EDELWEISS dark matter search experiment reported the measurements
of Germanium recoils in coincidence with muon signals in scintillators~\cite{edelweiss2}. 
The neutron-induced recoils have energies up to 60 keV, corresponding to neutron energies
up to GeV, depending on the scattering angle. The LVD experiment at Gran Sasso has also
reported neutron-induced recoil energy up to 300 MeV in liquid scintillators~\cite{lvd}.  
With bigger acceptance, a large liquid
scintillation detector holds promise to directly measure fast neutrons in a deep
underground environment.  
\par
The light response to nuclear recoils caused by neutrons within liquid 
scintillators are usually measured using the TOF technique
or the unfolding method. The latter one requires a response function
to unfold the visible energies ``seen" by PMTs to the recoil energy  
caused by incident neutrons. 
Such a function has been widely studied for neutron
energies from a few MeV to several hundred 
MeV~\cite{cecil, batch, verb, gul, aksoy,nakao} with small scintillation
detectors (several liters).   
Several Monte Carlo codes such as CECIL~\cite{cecil}, O5S~\cite{o5s} and 
SCINFUL~\cite{scinful} are developed to calculate the neutron response
function in liquid scintillators. 
\par
Aimed at characterizing fast neutrons at deep underground sites, 
a neutron detector has been fabricated at the University 
of South Dakota (USD).
It consists of an aluminum tube, one meter long and 5 inches in diameter,
filled with 12 liters EJ-301 liquid scintillator.  
Two 5-inch Hamamatsu PMTs (R4144) are attached to both ends of the tube through 
Pyrex windows to collect the scintillation light.
Detailed calibration procedures and neutron-gamma separation techniques are discussed
in Ref.~\cite{LSneu}. In this paper, the detector responses to 
atmospheric neutrons are studied. We show the measured nuclear recoils with energy up to 
$\sim$ 50 MeV using two years data
collected at the Soudan Mine.       
\section{Energy calibration for high energies}
\begin{figure*}[htp!!!]
\includegraphics[width=0.48\textwidth]{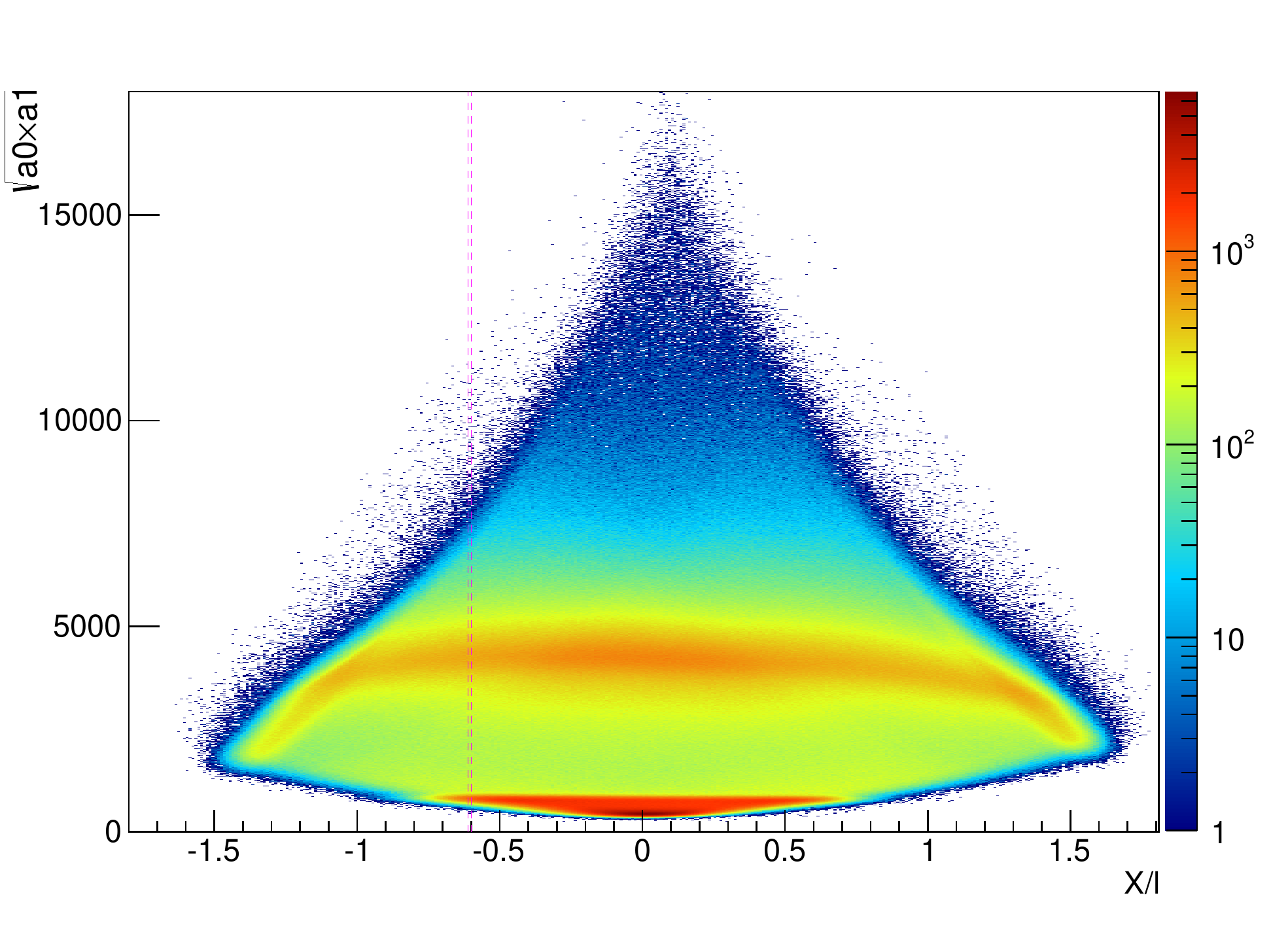}
\includegraphics[width=0.48\textwidth]{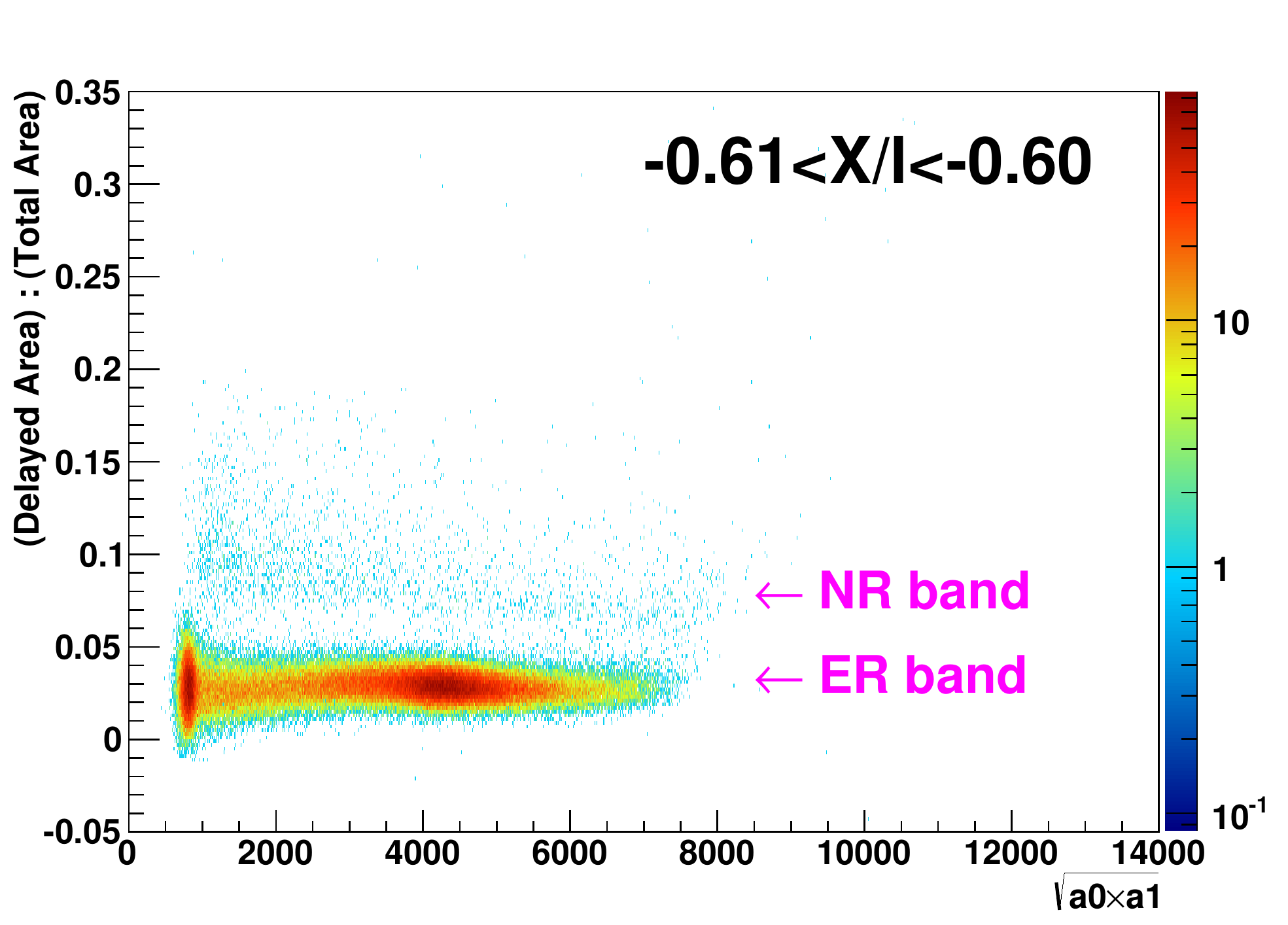}
\caption{\label{posiEn} The left plot is the detector response to scintillation light  
versus the position for a surface background run with a live time of 
19.4 days. The right plot shows the separation of nuclear recoils (NR) and 
electron recoils (ER) for a
position range $-0.61<X/l<-0.60$, where $X$ represents the distance of an energy deposition to
the middle of the detector and $l$ stands for the attenuation length in the scientillator. The position range is also marked 
as the dashed lines in the left plot.  
}
\end{figure*}

A background run with a live time of 19.4 days was conducted in a surface building
 at the USD campus prior to moving the detector underground. 
Following the calibration strategy we developed in Ref.~\cite{LSneu}, the 
detector responses to scintillation lights are shown in FIG.~\ref{posiEn}. 
An example of NR/ER discrimination from a very narrow position 
range is also shown in the right plot.   
\par
According to Ref.~\cite{birkslaw}, the light output $L(E)$ is a function of the stopping
power $dE/dx$ for a charged particle travelling in the scintillator
\begin{equation}\label{inte_birks}
 L(E) = S\int^{E}_{0}\frac{dE}{1+kB\frac{dE}{dx}},
\end{equation} 
where $S$ stands for the scintillation efficiency and $kB$ is called Birks' 
constant for the specified medium. As can be seen in Eq.~(\ref{inte_birks}), for electrons with energies greater
than $125$ keV in the scintillator, the stopping power becomes very small~\cite{birkslaw} which
makes $kB\cdot(dE/dx)\ll 1 $, as a result, the light output $L(E)$ can be simplified to be
a linear relation to higher energies.  
\par
This liquid scintillation detector is calibrated from 1 MeV to 20 MeV 
by using $^{22}$Na (1.275 MeV), 
AmBe sources (4.4 MeV), and the minimum ionization peak from cosmic muons (20.4 MeV). 
Applying the position independent variable $\sqrt{a0\times a1}$, where 
$a0$ stands for the total charge converted from PMT0 and $a1$ is the total
charge converted from PMT1,
a second order polynomial function is assumed to fit the calibration
curve for energy below 20 MeV. For energies above 20 MeV,
the first order approximation is a simple extension from the low energy
calibration curve. 
Considering the background signals with the energy above 10 MeV are dominated by
the well-understood surface muons, the calibration curve above 20 MeV is then 
adjusted by the detector response to the surface muons.  
A GEANT4~\cite{geant4} (GEANT4.9.5.p02 + Shielding module physics list) based
simulation is conducted with a modified Gaisser's formula~\cite{modGaisser} 
(sea level)
to sample the shape of the energy spectrum and angular distribution of input muons.
The simulated detector response to muons is compared with data by assuming 
a linear relation, justified by Eq.~(\ref{inte_birks}), between the light output $\sqrt{a0\times a1}$ and the energy 
deposition ($>20$ MeV) in the detector.   
The slope of the calibration line ($>20$ MeV) is then determined by fitting 
the detector response to the simulated muons with the corresponding data.
\begin{figure*}[htp!!!]
\includegraphics[width=0.96\textwidth]{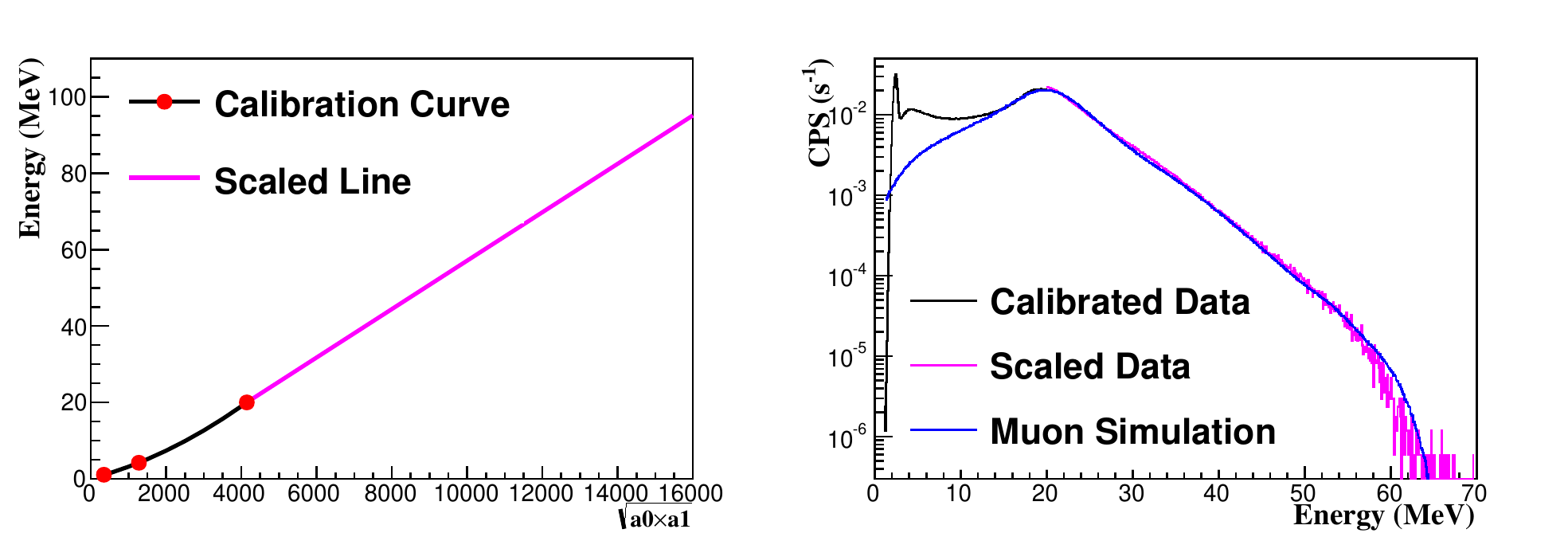}
\caption{\label{cali} The left plot is the calibration curve
of the light output versus the energy deposition in the detector. 
The calibration curve ($<20$ MeV) is fitted by the three calibration sources. 
The scaled line ($>20$ MeV) is obtained by the simulation.
The right plot is a comparison of the detector response to muons between 
the simulation and data. The black line represents the calibration data. The magenta
line stands for the scaled data from the simulation. The blue line is the muon simulation.}
\end{figure*}

The plots in FIG.~\ref{cali} demonstrate how the high energy calibration 
line (left) is determined by fitting data with the simulation (right).
For the lower energy range ($<10$ MeV), the data are overwhelmed by 
the internal contamination and environmental gamma rays, which explains why the
data and the muon simulation do not match at such range.
It is worth mentioning that the absolute normalization of the detected muon
intensity is 16.6\% higher than the flux from sea level. This is 
reasonable because the surface muon data was taken at 
the campus of the University
of South Dakota with an elevation of 1221 feet~\cite{elevation} 
above the sea level.  
\par 
\begin{figure}[htp!!!]
\includegraphics[width=0.48\textwidth]{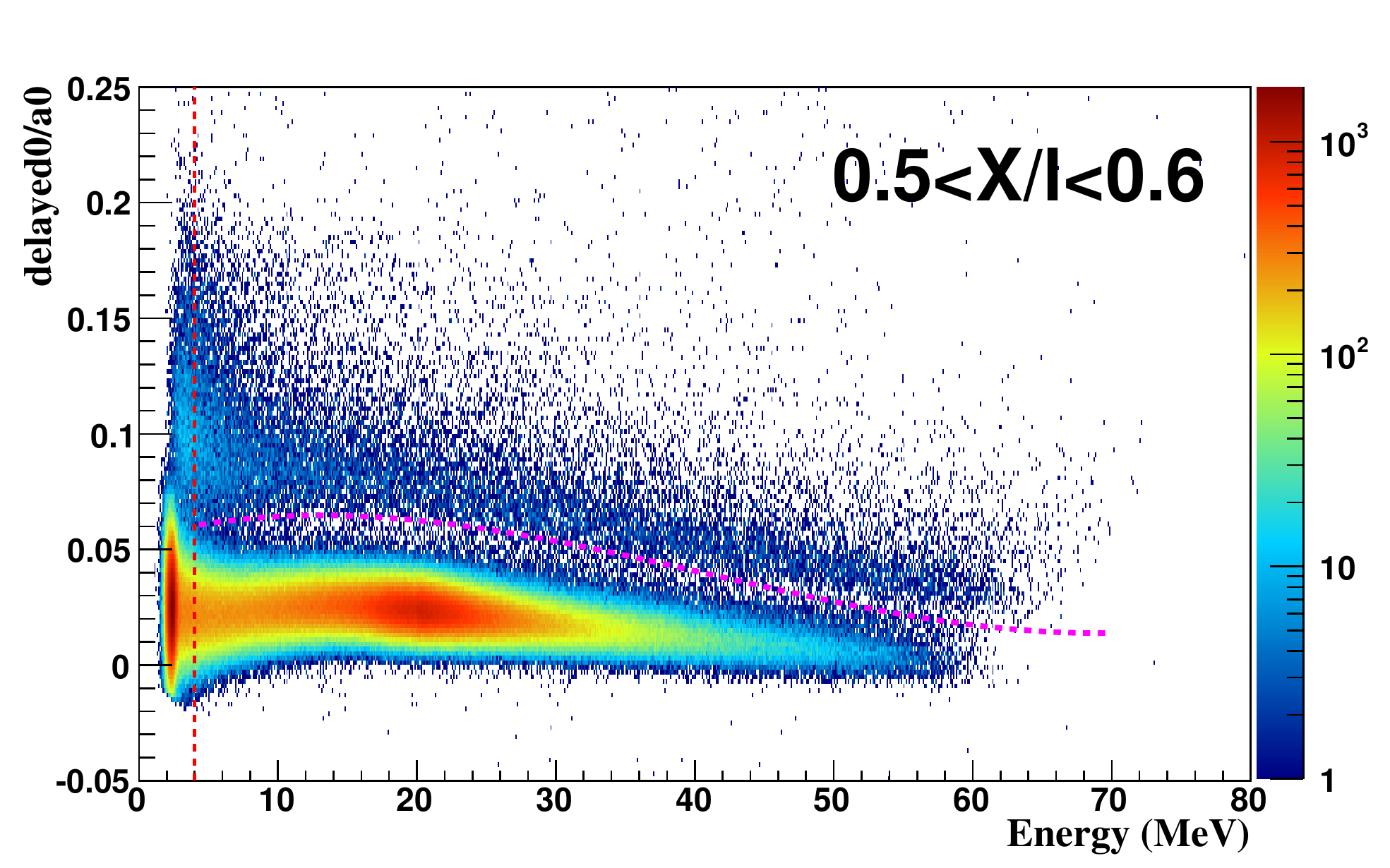}
\caption{\label{ngSeparation} 
The plot shows the events with positions between 0.5$<$$\frac{X}{l}$$<$0.6 (see the left plot in Fig.~\ref{posiEn}) from a surface
background run with a live time of 19.4 days.
The vertical line denotes the energy threshold setting at 4 MeV. The 
fitted separation curve is a three-order polynomial function with
the parameters (0.0550676, 0.00157856, -7.17262e-05, 5.82431e-07). 
}
\end{figure}

\begin{figure*}[htp!!!]
\includegraphics[width=0.98\textwidth]{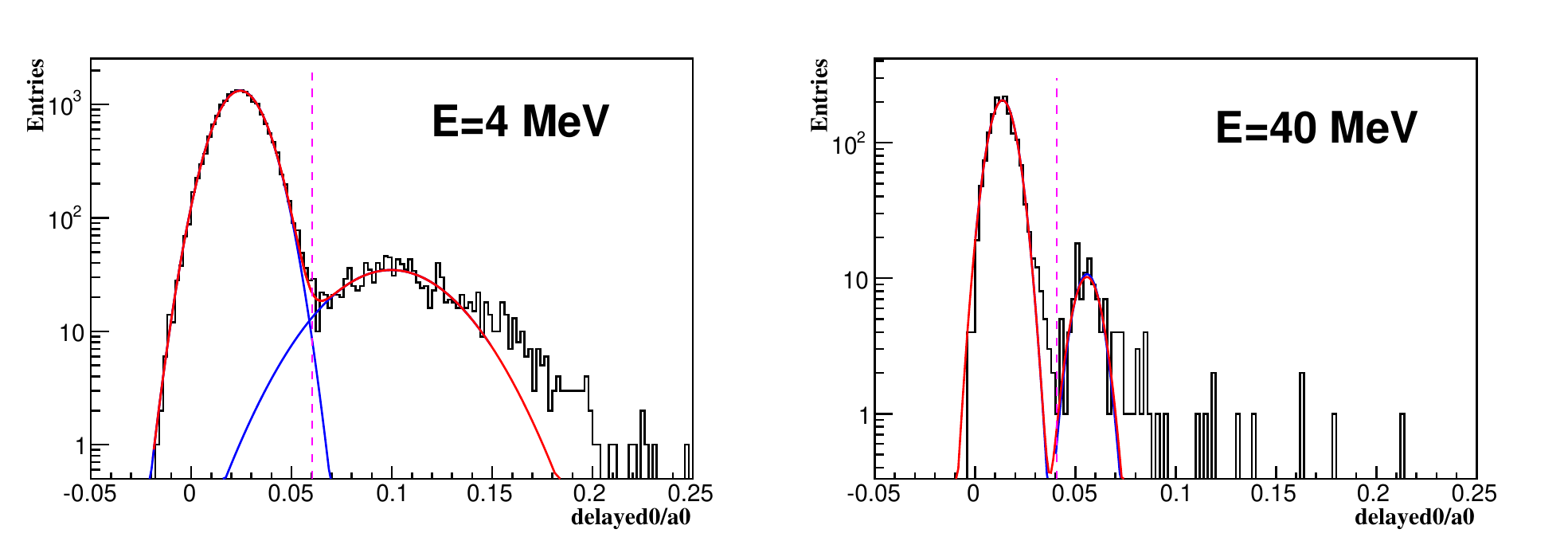}
\caption{\label{ngEfficiency} 
The separation of neutrons and gamma rays at the energies of 4 MeV (left) and 40 MeV (right) are shown. 
For the energy of 4 MeV, the gamma-ray contamination is 1.10\% and the acceptance of neutrons is 91.73\%.
For the energy of 40 MeV, the gamma-ray contamination is 0.01\% and the acceptance of neutrons is 99.19\%.
}
\end{figure*}
The plots in FIG.~\ref{ngSeparation} and FIG.~\ref{ngEfficiency}, as an example,  represent the selected 
gamma rays and neutrons to reconstruct the surface muon and neutron energy spectrum.     
Utilizing the calibrated energy scale, we can assume that the linear relation between the energy deposition and the 
light output is the same for electron recoils
and nuclear recoils after the correction for quenching.    
\section{Light output response to fast neutrons}
\begin{figure}[htp!!!]
\includegraphics[width=0.48\textwidth]{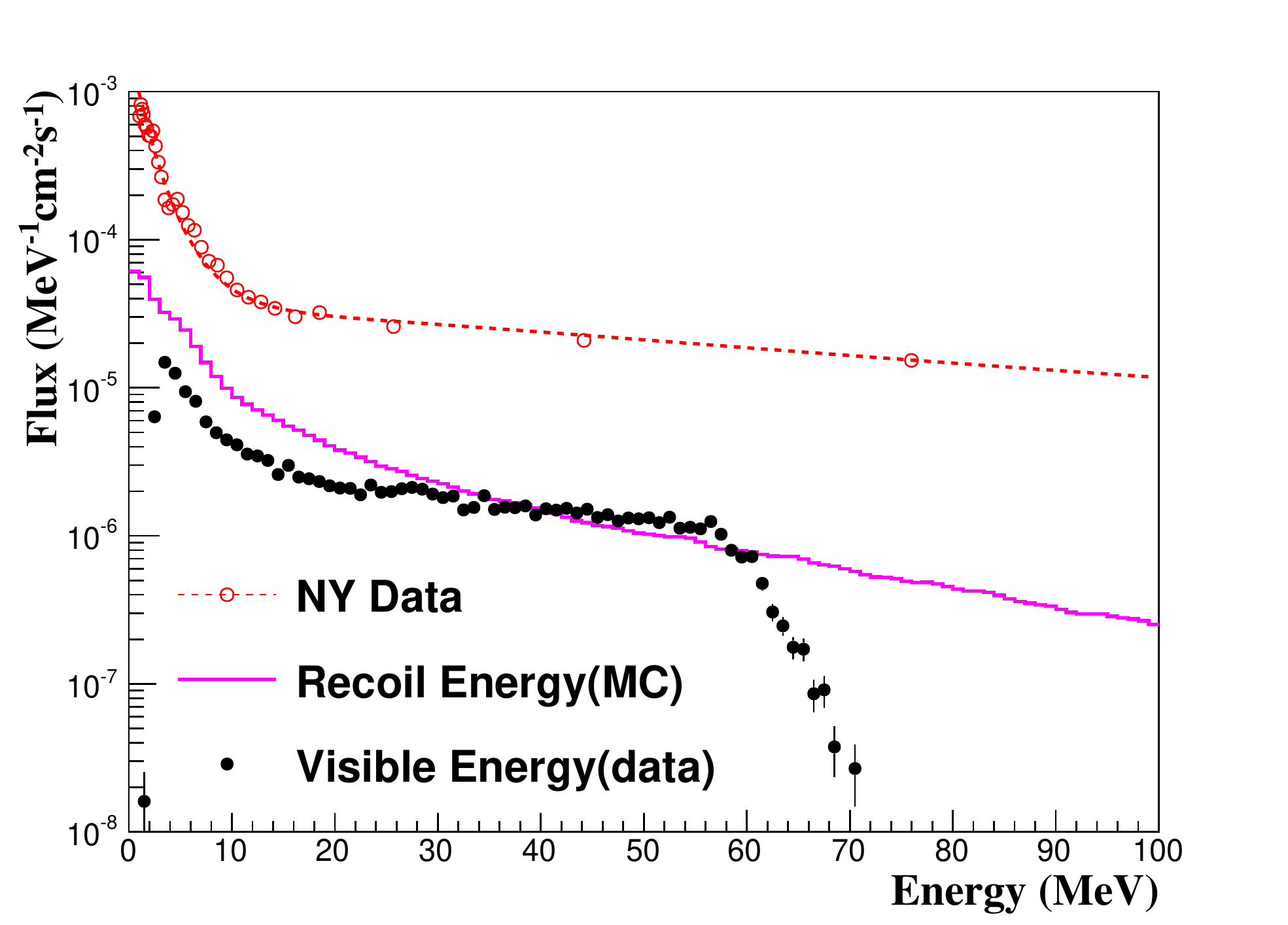}
\caption{\label{compRecoil} The detected visible energy (solid dots) compared with
the simulated recoil energy (solid line). The input neutron energy spectrum from
the surface measurement in New York city is also listed (dashed line with open dots). 
}
\end{figure}

The results from the surface background run in FIG.~\ref{ngSeparation} show
the separation of nuclear recoils from electronic recoils. After selecting only those events from the NR band,   
the visible energy from 
nuclear recoils in the detector is shown in FIG.~\ref{compRecoil} (solid dots).  
The cut-off at $\sim$2 MeV is caused by energy threshold set on the trigger while 
the waterfall at 60-70 MeV is caused by the saturation of the Analog-to-Digital
Converter (ADC) channels. 
A simulation is performed by adopting the neutron energy spectrum from 
Ref.~\cite{neuNYdata} as an input. Although it is a neutron measurement in New York city, 
it can serve as a reasonable approximation by taking just
the shape of the neutron energy spectrum with the intensity to be determined later by
comparing the simulated detector response to the experimental data. 
\par
\begin{figure}[htp!!!]
\includegraphics[width=0.49\textwidth]{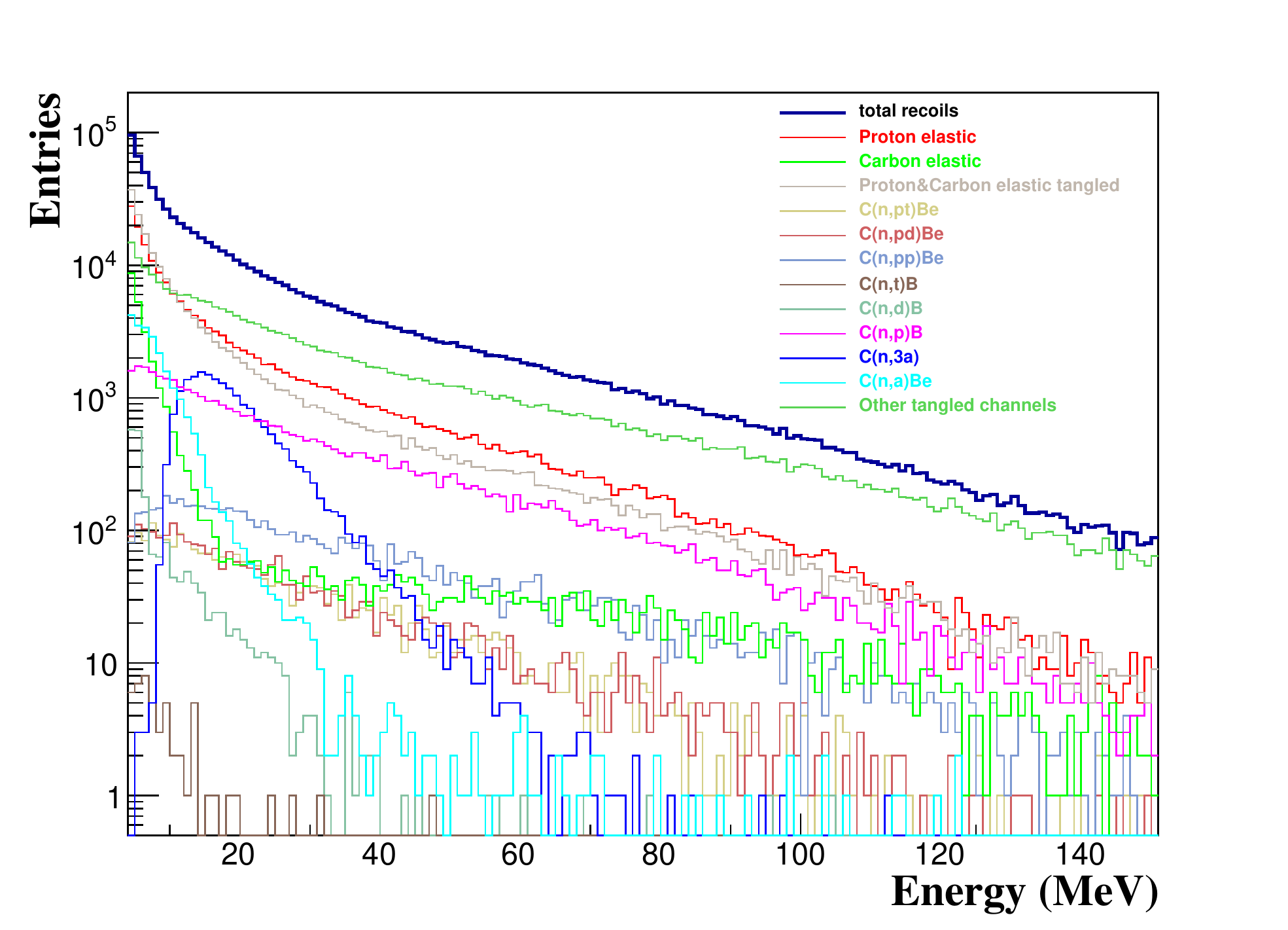}
\caption{\label{recoilChannels} The number of interaction types in the
nuclear recoil events. The total recoil energy is defined 
as a sum of all energy deposition
of the event which contains at least one ion scattering in the scintillator. 
This is a simulation result by adopting the incident neutron flux from 
Ref.~\cite{neuNYdata}. The energy threshold of the input neutrons is set to be 4 MeV. 
}
\end{figure}
\begin{figure}[htp!!!]
\includegraphics[width=0.48\textwidth]{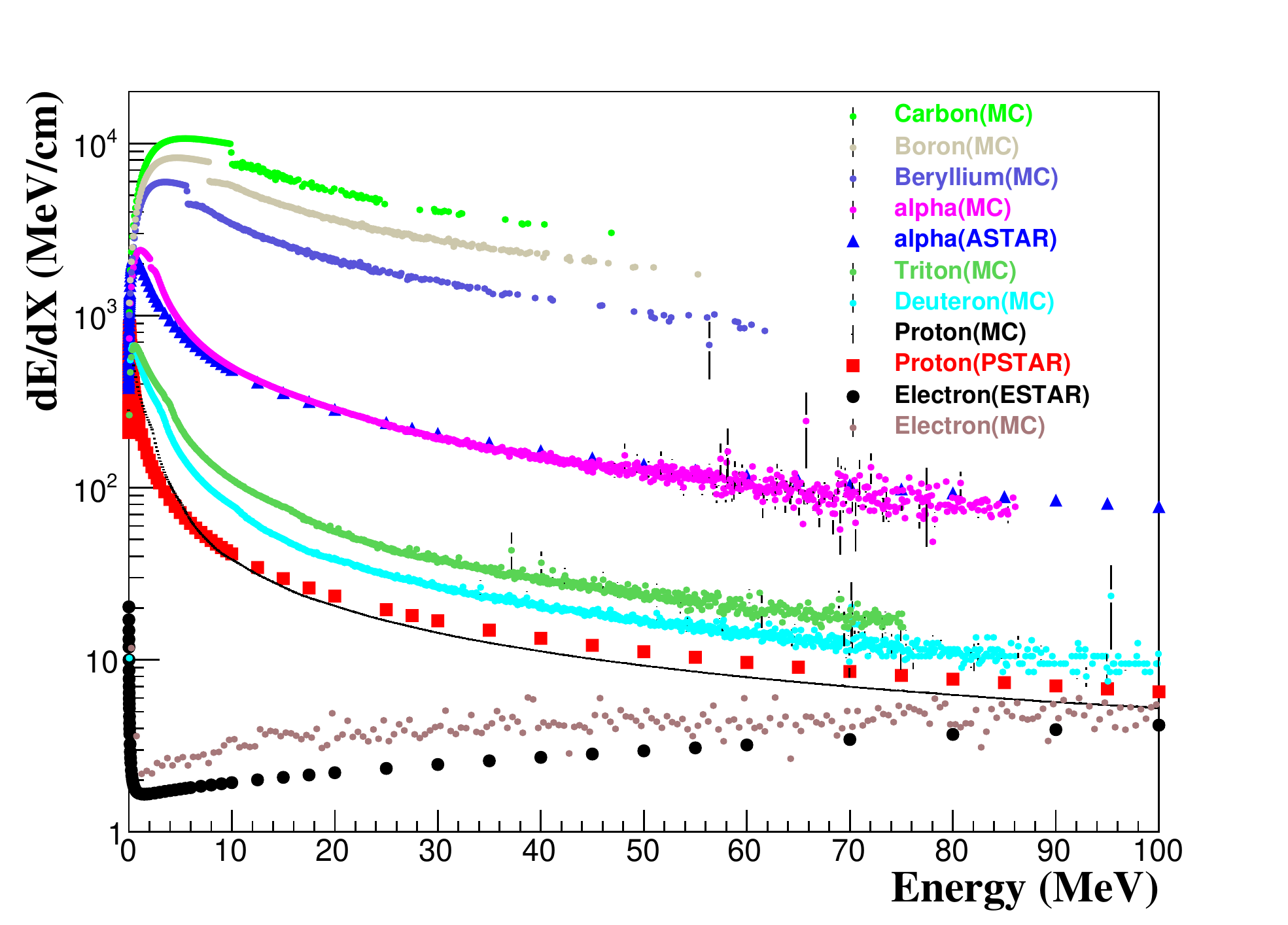}
\caption{\label{dedx} Stopping power obtained from simulations for carbon,
boron, beryllium, alpha, triton, deuteron, proton and electron in the EJ-301 scintillator. As
a comparison, the stopping power from ESTAR~\cite{estar}, PSTAR~\cite{pstar} and 
ASTAR~\cite{astar} for electron, proton and alpha in plastic scintillators 
are also shown, respectively. The density of liquid scintillator (0.87 $g/cm^{3}$) is applied 
to the mass stopping power of plastic scintillators in order to convert the unit $MeV\cdot cm^{2}/g$
to $MeV/cm$.     
}
\end{figure}

The nuclear recoils from neutrons in the detector are simulated and the result is shown 
in FIG.~\ref{compRecoil} (solid line). No cuts are applied to the simulated curve yet since 
we need to understand the light output response from nuclear recoils. 
The measured visible energy to nuclear
recoils is also shown in FIG.~\ref{compRecoil} (black dots). 
A quenching factor matrix 
exists between the recoil energy 
and the visible energy ``seen" by the PMTs. 
The light output in the liquid scintillator can be described by 
Birks' relation~\cite{birkslaw, birks}:
\begin{equation}\label{diff_birks}
 \frac{dL}{dx} = \frac{S\frac{dE}{dx}}{1+kB\frac{dE}{dx}},
\end{equation}
where $\frac{dL}{dx}$ represents the light output per unit path length.
The quenching
factor for nuclear recoils is defined as the ratio of light yield of 
ions to that of electrons of the same energy~\cite{quenchingfactor}. This definition allows us to 
calculate the quenching factor by rewriting
Eq.~(\ref{diff_birks}) as:
\begin{equation}\label{quenching_birks}
 Q_{i}=\frac{L_{i}(E)}{L_{e}(E)} = 
	\frac{\int^{E}_{0}\frac{dE}{1+kB(\frac{dE}{dx})_{i}}}
		{\int^{E}_{0}\frac{dE}{1+kB(\frac{dE}{dx})_{e}}}.
\end{equation}
The Birks' constant $kB$ is believed to be the same for all particles
in the same medium. For the liquid scintillator EJ-301, it has been measured to 
be $\sim$ 161 $\mu$m/MeV~\cite{ej301qf}. The interactions of a fast neutron in
the liquid scintillator are dominated by multiple scattering processes and 
therefore generate multiple ion recoils (see FIG.~\ref{recoilChannels}). 
This means that even if the total recoil
energy is the same, it could be composed of single
or multiple ion recoils which have different quenching factors in the scintillator.  
Theoretically, a combined quenching factor could be calculated if we track all 
scattering processes in simulations.   
Other than the Birks' constant, the other remaining variable is the
stopping power for each ion in the scintillator. 
FIG.~\ref{dedx} summarizes the $dE/dx$ functions obtained from the simulations. 
The stopping power converted from the NIST web database ESTAR, PSTAR and
ASTAR for electrons, protons and alphas in scintillators
are also listed, respectively\cite{estar, pstar, astar}. Within the web database, the material most similar 
to EJ-301 
is the plastic scintillator. Therefore we take the mass stopping power of 
the plastic scintillator and convert it to be the stopping power for the liquid scintillator 
simply by applying 
the density of EJ-301 scintillator. The comparison in FIG.~\ref{dedx} 
shows that the stopping power function  
for alpha (ASTAR) and proton (PSTAR) have a reasonable agreement with the calculations from
GEANT4 simulation. The stopping power of 
electrons from GEANT4 is about 20\% higher than that from the ESTAR.        
This discrepancy is likely caused by the lack of correction for shell-effect,  in GEANT4 for electrons. 
The stopping power we used to calculate the quenching factors are all from GEANT4 simulations. Therefore, this discrepancy
is taken into account as part of the total uncertainty in deriving the measured neutron flux in section IV-B.

\par
\begin{figure}[htp!!!]
\includegraphics[width=0.48\textwidth]{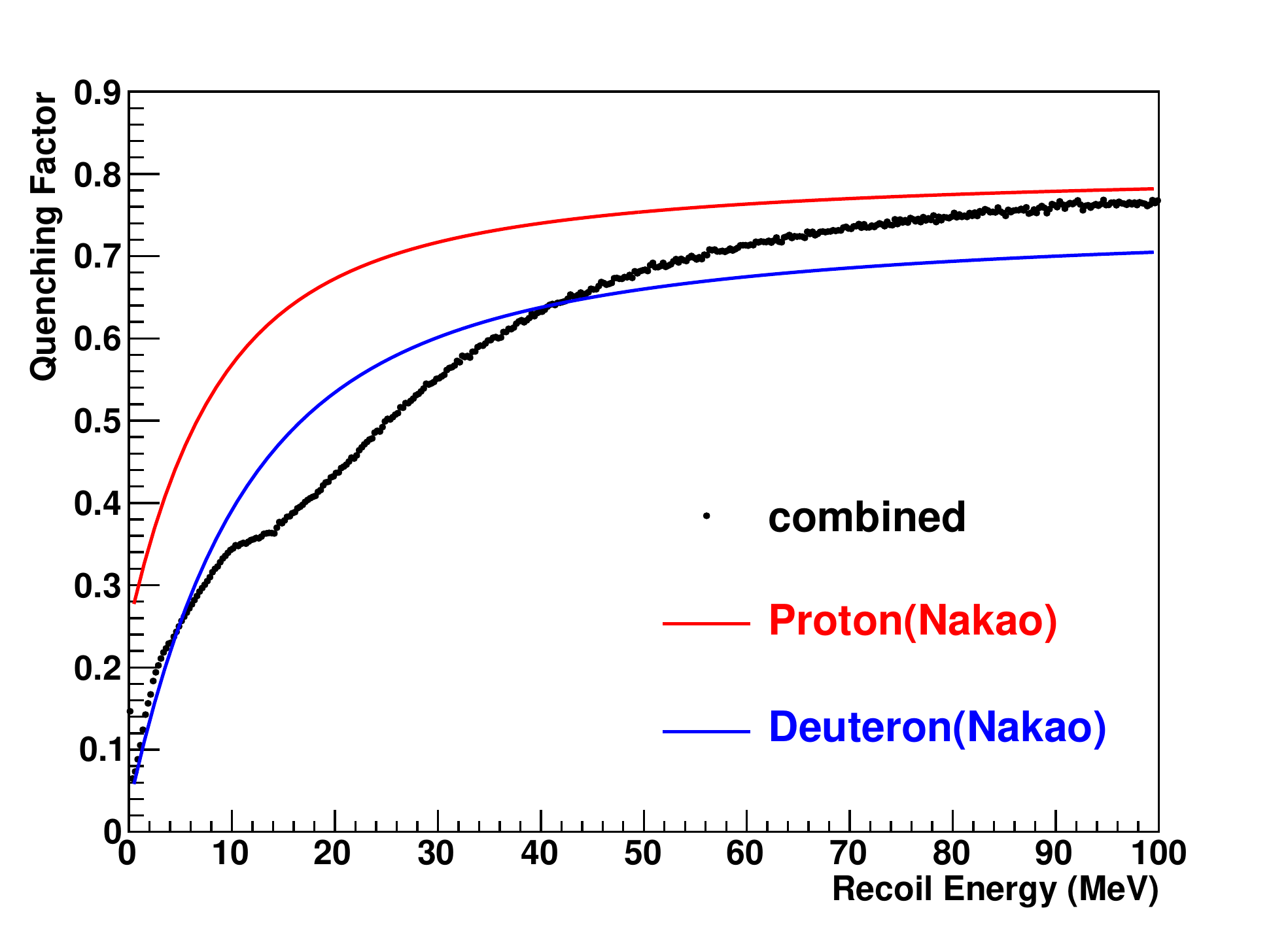}
\caption{\label{quenching} The combined quenching factor for EJ-301 scintillator 
calculated from simulations (black dots). 
The parameterized quenching function for proton (red line) and 
deuteron (blue line) in EJ-301 scintillator~\cite{nakao} are also
plotted.   
}
\end{figure}

By applying Eq.(\ref{quenching_birks}) step by step for each 
nuclear recoil event in the simulation,
a combined quenching factor function is generated as shown in FIG.~\ref{quenching}.
For the recoils with energy in the range of MeV, carbon scattering plays an important role in terms of quenching effect 
which drives the quenching factor much lower when compared to that of the proton only. 
When neutron energies approach $\sim 13$ MeV, $C(n, 3\alpha)$ processes start to build 
up. The combined quenching factor of multiple $\alpha$s has an even lower quenching factor
when compared to single $\alpha$ with the same total recoil energy in the scintillator.
This explains why there is a ``knee" at such an energy range. For high energy recoils, 
the most of the energy is taken by proton scatterings which make the combined quenching 
effect approaching to that of proton recoil in this range.     
\par
\begin{figure}[htp!!!]        
\includegraphics[width=0.48\textwidth]{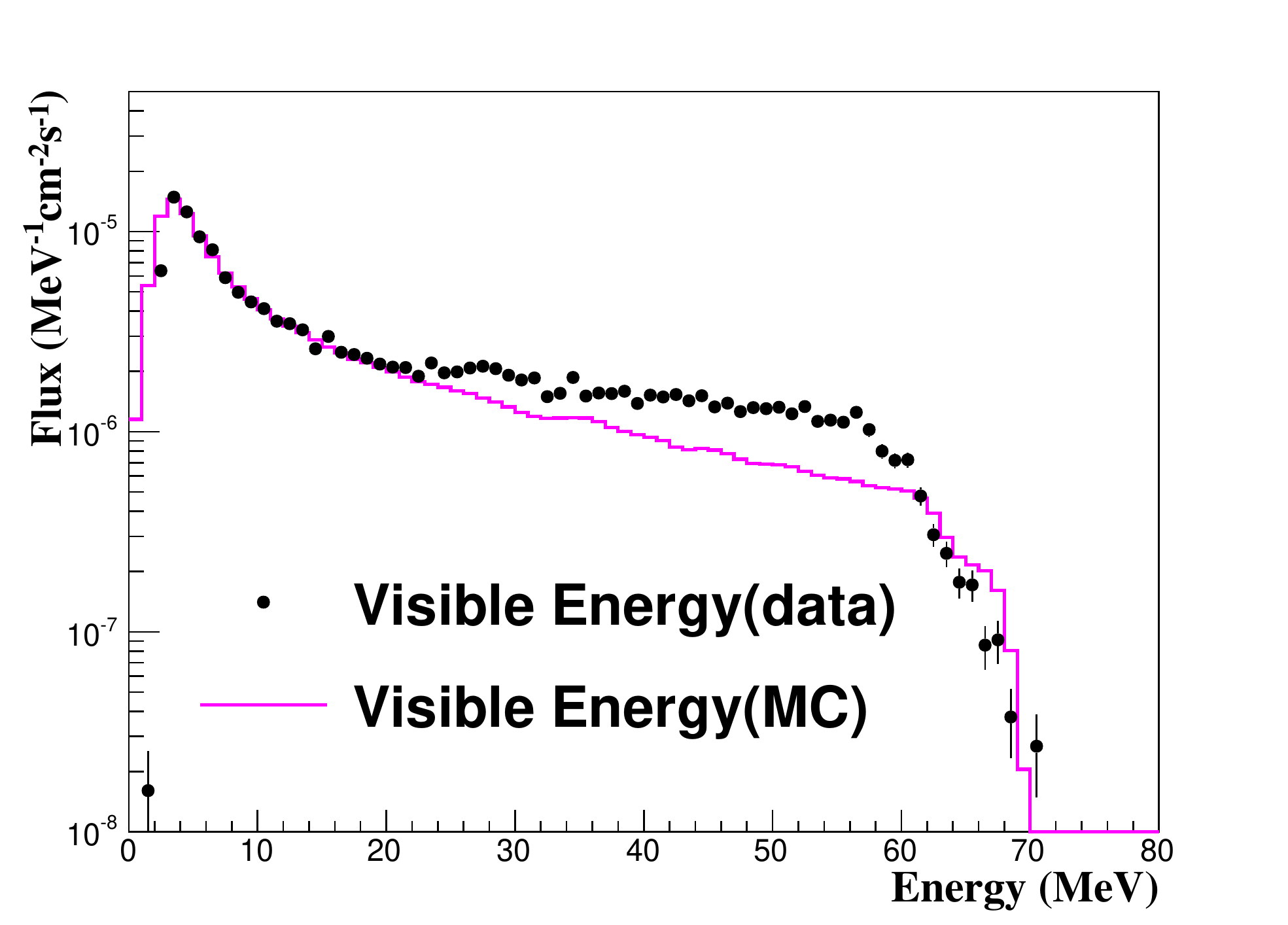}
\caption{\label{compDataWithMC}
Comparison between the measured data (solid dots) and the simulation
(solid line) for the visible energy of nuclear recoils.
}
\end{figure}

 Applying the calculated quenching factor to each ion recoil in the simulation, the simulated 
visible energy is obtained and compared with the data (see FIG.~\ref{compDataWithMC}).  
The simulated results show a good agreement with data for the energies below 20 MeV.
For the energies above 20 MeV, the current simulation yields less nuclear recoil events by 
a factor of $\sim$2 than that of 
data. There are two reasons that may cause such a discrepancy.
One is the limited understanding of the detector response to high energy neutrons, such as energy and 
position reconstructions. The other is the input source in which we only implemented neutrons in the simulation that 
ignored nuclear recoils induced by other possible sources such as protons, alphas, and high-energy gammas.      
\section{Muon and neutron measurements at Soudan Mine}
The Soudan underground laboratory is located at a 690 m deep (2100 m.w.e) underground
facility  in the Soudan Mine, Minnesota. Several underground experiments such as MINOS~\cite{minos}, CDMS~\cite{cdms0}, and 
CoGeNT~\cite{cogent0} are running there.  
With such a large rock overburden, the background from cosmic-ray muons is dramatically 
suppressed. The muon flux, passing through a horizontal surface, at Soudan Mine was measured to be $1.77\times10^{-7}$cm$^{-2}$s$^{-1}$
from the MINOS far detector~\cite{minos1}.
The neutron flux in the laboratory is dominated by the radioactivity
from the surrounding rocks through $(\alpha, n)$ reactions and fission decays which have the
most of energies below 10 MeV. For those neutrons with higher energies, 
the production comes primarily from cosmic-ray muons through spallation processes. 
\subsection{Simulation of muons and the muon-induced secondaries}
\begin{figure*}[htp!!!]
\includegraphics[width=0.49\textwidth]{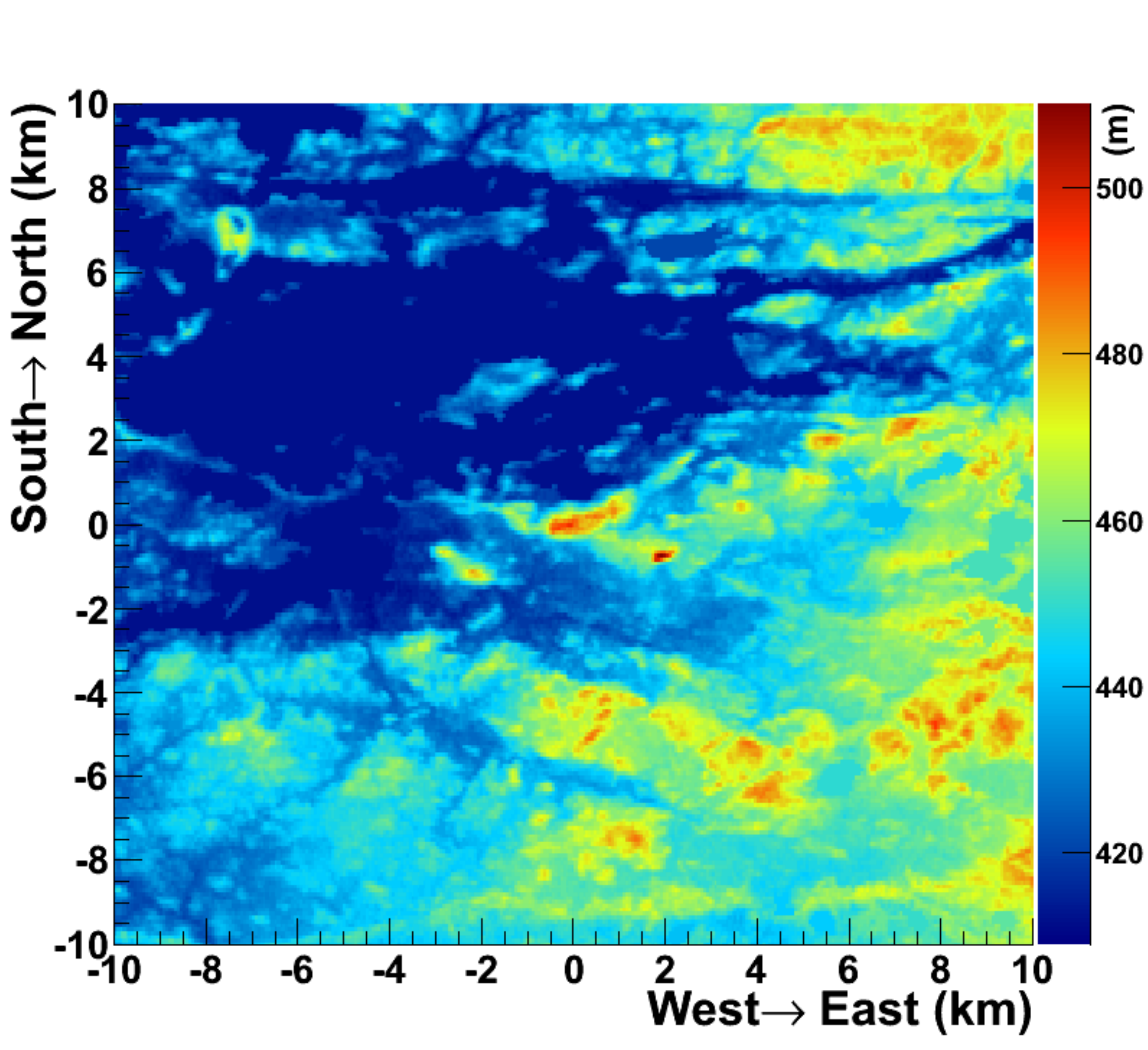}
\includegraphics[width=0.49\textwidth]{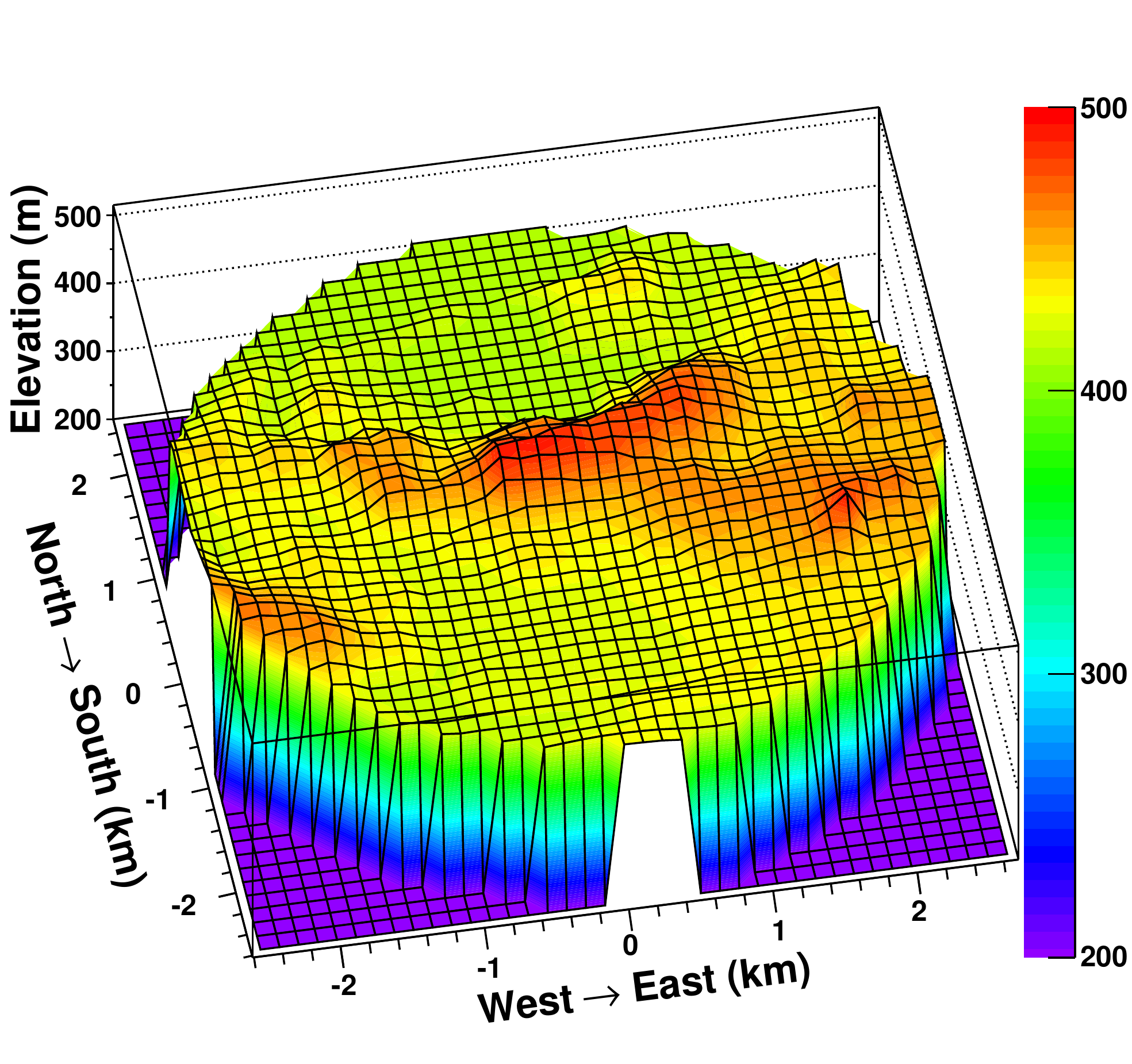}
\caption{\label{map}
Shown is a digitized elevation map around the Soudan Mine area from the satellite data~\cite{usgs}. 
The total extension of 20 km $\times$ 20 km is used in the simulation (left plot). 
The right plot is a zoomed-in view of the central area.  
}
\end{figure*}

\begin{figure*}[htp!!!]
\includegraphics[width=0.98\textwidth]{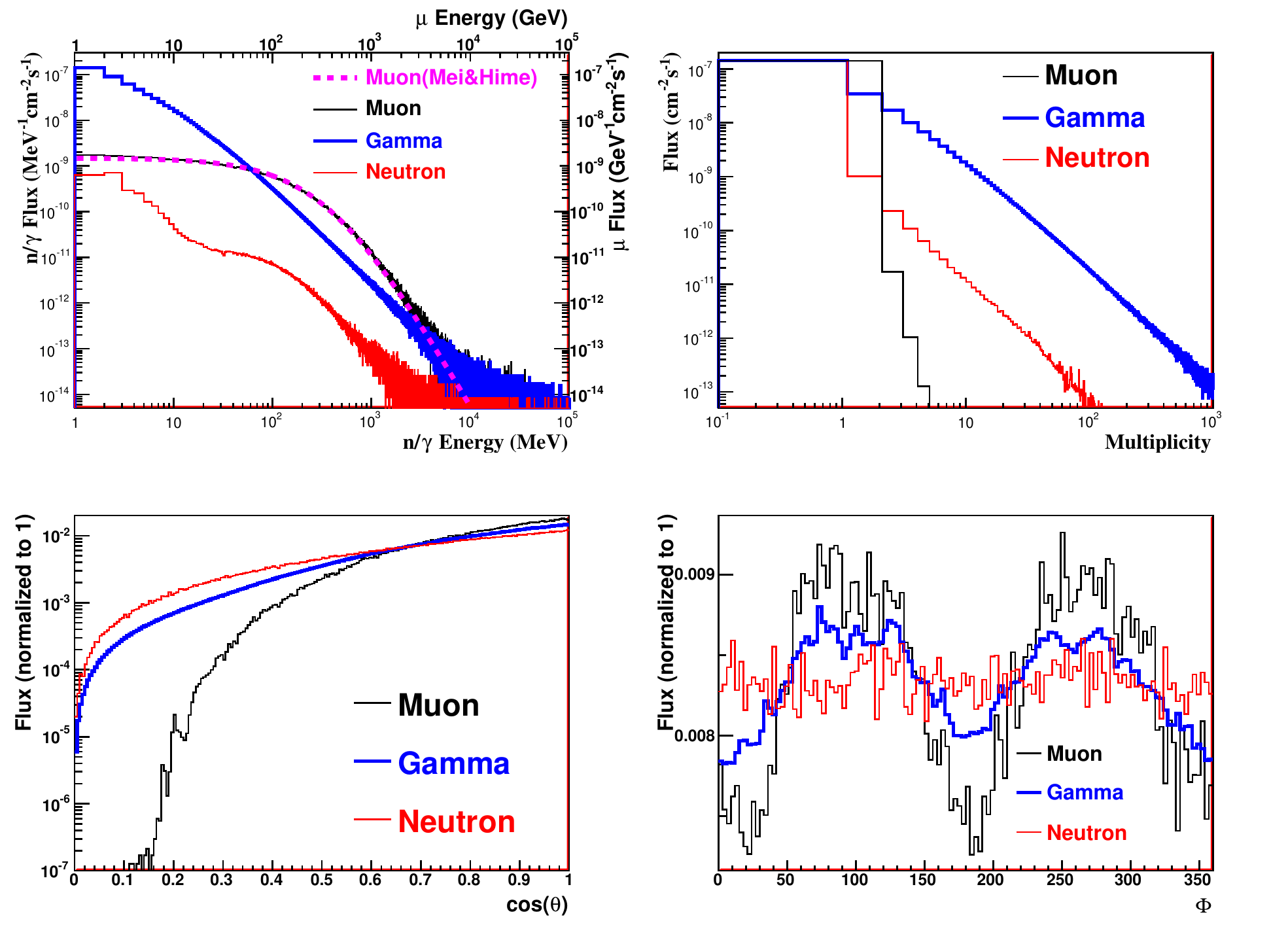}
\caption{\label{soudan_sim}The energy and angular distribution of the muons and the associated neutrons, gamma rays in the cavern at the depth of Soudan 
 from the Monte Carlo simulation. The energy threshold set in the simulation 
is 1 GeV for muons and 1 MeV for neutrons/gammas.  
Note that the energy scale of muons is in GeV while that of neutrons and gamma rays are in MeV. 
The dashed line shown in the top-left plot is the muon energy spectrum predicted by Mei\&Hime~\cite{mei} at Soudan depth.
$\theta$ is defined as the zenith angle to the vertical direction (downwards).
Azimuthal angle $\Phi$ is defined as an observation angle to the east direction.
$\Phi=0$ stands for particles coming from the east while $\Phi=90$ represents particles
coming from the north.  
}
\end{figure*}

The high-energy particles produced underground are induced by
cosmic-ray muons which penetrate from the Earth's surface down to the mine. 
The intensity of these muons varies from one location to another, depending on the 
altitude, and profile of the mountains on the surface,  as well as the rock densities
along the path of muons.  
In order to quantify those cosmogenic events, 
a full GEANT4-based simulation has been conducted by adopting the surface mountain profile
from the CGIAR satellite data~\cite{usgs} with the extension of 20 km $\times$ 20 km 
as shown in FIG.~\ref{map}. Note that a PeV muon has a travel distance
6$\sim$7 km in rocks on average. A 10 km radius would be sufficient to serve our simulation purpose.
According to the information provided by
MINOS experiment, its far detector situates at (longitude: 92$^{o}$14'28.51443''W, 
latitude: 47$^{o}$49'13.25409''N)~\cite{location}. 
A 20 $m^{3}$ cavern is assumed with the center located at
(0, 0, -217 m) in the map.    
A typical rock composition of Ely Greenstone is selected with an average rock density
of 2.85 $g/cm^{3}$~\cite{soudanRock} in the simulation. Muons sampled from the modified 
Gaisser's formula~\cite{modGaisser} are then tracked from the surface of the mountains down
to the cavern. Finally, muons and the associated secondaries are collected at the ceiling 
and walls of the cavern with their
energy and angular distribution displayed in FIG.~\ref{soudan_sim}.   
\par
The top-left plot in FIG.~\ref{soudan_sim} compares the energy spectrum of the simulated muons in the cavern with the 
associated secondary neutrons and gamma-rays from surrounding rocks. 
The energy scale of the muons is in GeV while the neutrons and gamma-rays are in MeV.
For the input surface muons, only single muons are simulated (no bundles). Therefore, the events with multiple
muons in the results are due to pion decay on-flight.  
The comparison of their multiplicities in the top-right plot reveals the muon
shower information:
\begin{itemize} 
\item{The relative ratios for the production of neutrons and high-energy gamma rays are
 counted to be $N_{\mu}(E_{\mu}$$ >$ 1 GeV)$:N_{n}(E_{n}$$ >$ 1 MeV)$:N_{\gamma}(E_{\gamma}$$ >$ 1 MeV) =$1:0.0115:0.715$, 
which means that there are 1.15\% of 
primary muons (E$_{\mu}$$>$ 1 GeV)
generated neutrons above 1 MeV and 71.5\% of primary muons  (E$_{\mu}$$>$ 1 GeV) generated gamma rays above
1 MeV  at the depth of Soudan.}
 \item{The average multiplicity per muon-induced event is $M_{\mu}:M_{n}:M_{\gamma}=1.0:2.6:9.9$, which means that the 
average multiplicity for neutrons is about 2.6 and for gamma rays is about 9.9.}
\item{The angular distribution in the bottom plots indicate the angular correlation between
the primary muon and its secondaries. Comparing to the neutrons, the gamma rays are more peak-forwarded with respect 
to primary muons. The angular distribution of neutrons show a little correlation with respect to primary muons angular distribution.}
\end{itemize}
\par
The reconstructed neutron energy spectrum from the Monte Carlo simulation can be expressed using an analytic model suggested 
in Ref.~\cite{neuNYdata}:
\begin{equation}\label{analyticModel}
 \frac{d\Phi(E)}{dE} = \sum\limits_{j=1}^{2} c_{j}\exp\left[-\beta_{j}(\ln(E))^{2} + \gamma_{j}\ln(E)\right], 
\end{equation}
where the values of  $c_{j}$, $\beta_{j}$, and $\gamma_{j}$ are listed in Table~\ref{parameters}. 
Note that $c_{j}$ is a normalization factor which is determined by fitting the neutron 
energy spectrum using Eq.~\ref{analyticModel}, 
$\beta_{j}$ and $\gamma_{j}$ are the parameters suggested in Ref.~\cite{neuNYdata}. 
\begin{table}
\caption{Parameters for the analytic model in Eq.~\ref{analyticModel}}.
\begin{tabular}{c|ccc}
\hline
$j$ & $\beta_{j}$ & $\gamma_{j}$ & $c_{j}$ \\
\hline
1	&0.3500	&2.1451		&4.6283e-13\\
2	&0.4106	&-0.6670	&1.0097e-09\\
\hline
\end{tabular}
\label{parameters}
\end{table}

\begin{figure}[htp!!!]
\includegraphics[width=0.48\textwidth]{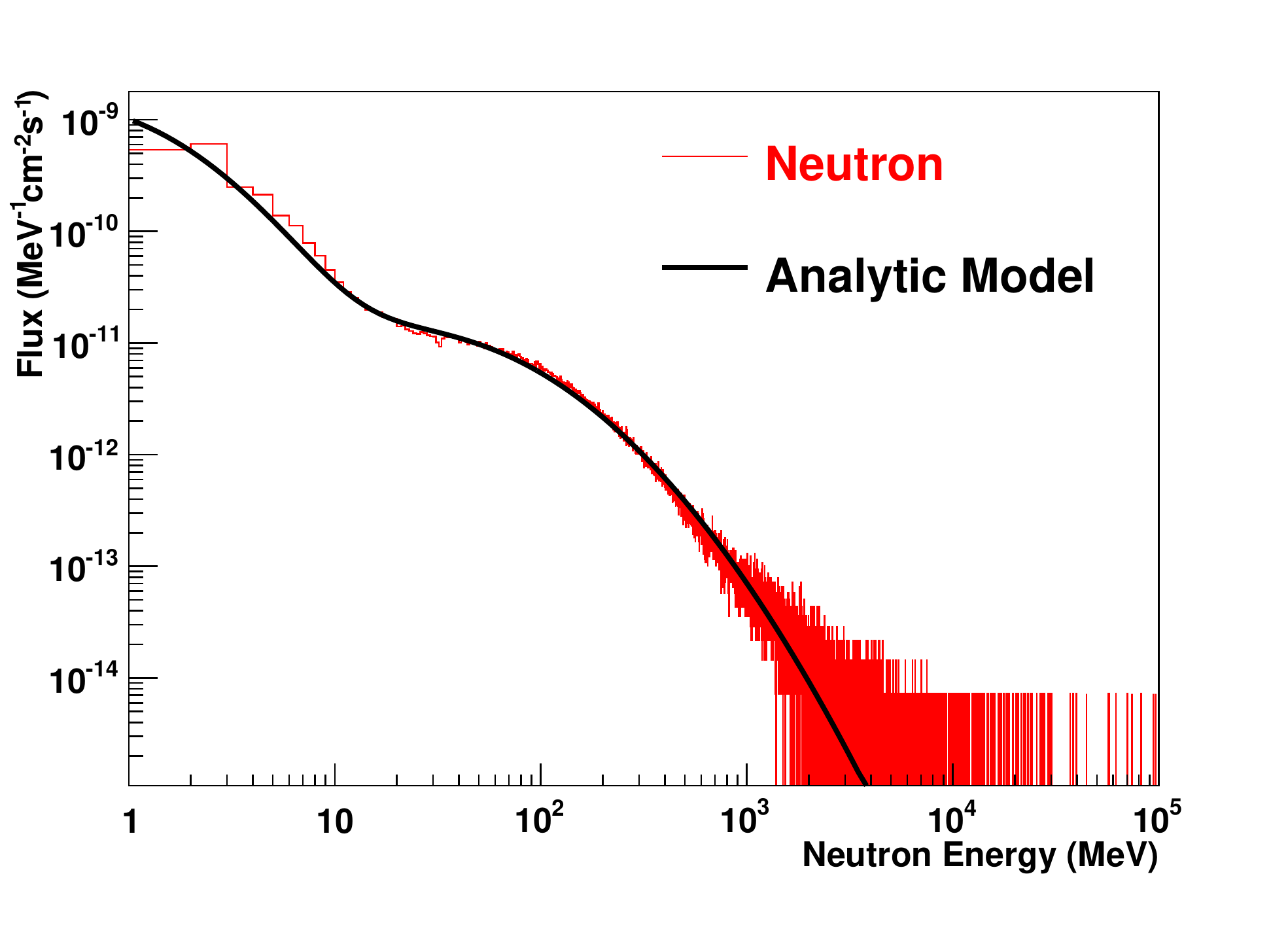}
\caption{\label{model} 
Reconstructed neutron energy spectrum at the depth of Soudan Mine (the red line).
The black line is the fitted analytic function~\cite{neuNYdata}.  
}
\end{figure}
\par
By normalizing the live time to that of the input muons at the surface level, 
the absolute fluxes for the muons, neutrons and gamma rays in the cavern are obtained to be 
$1.99\times10^{-7}$cm$^{-2}$s$^{-1}$($E_{\mu}$ $>$1 GeV), $5.72\times10^{-9}$cm$^{-2}$s$^{-1}$($E_{n}$ $>$1 MeV), 
and $8.57\times10^{-7}$cm$^{-2}$s$^{-1}$($E_{\gamma}$ $>$ 1 MeV),
 respectively. It is worth mentioning that the muon flux is defined as the muons passing through
 a horizontal surface where the muon flux is scaled by cos($\theta$).
\par
\begin{figure}[htp!!!]
\includegraphics[width=0.48\textwidth]{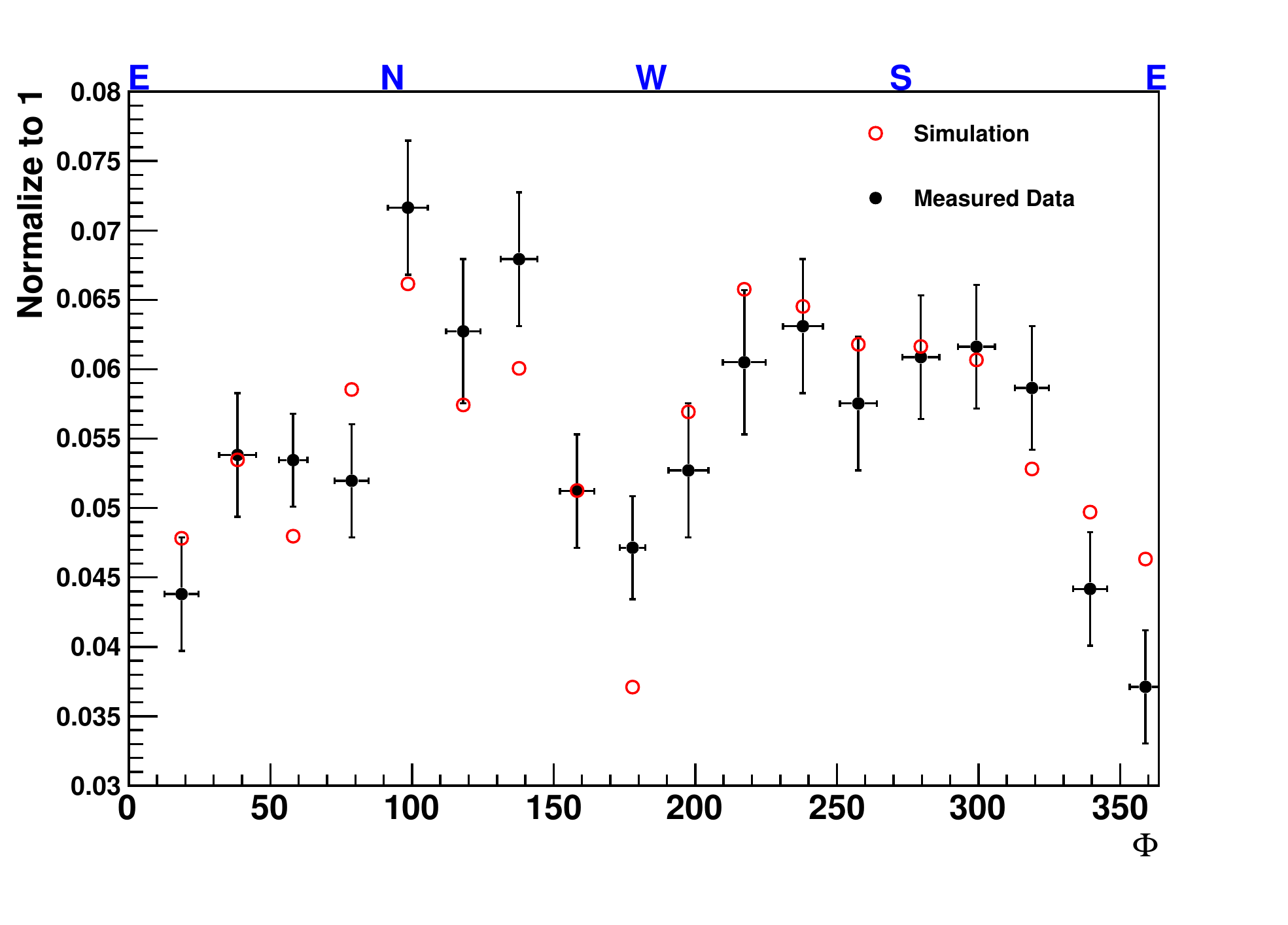}
\caption{\label{compPhi}
The comparison of muon azimuthal angle distribution at Soudan Mine.
The solid dots are from the experimental data measured using an active
muon veto~\cite{measuredphi}. The empty dots are from the simulation
result. The total fluxes are both normalized to 1. 
The orientation of muons from the East, North, West and South are denoted at the top.
}
\end{figure}

Although there are uncertainties from the input muon intensity and the rock density variations, 
the simulated fluxes serve as a first order of approximation for the intensity of 
muons and the induced secondaries in the cavern. The absolute normalization of the flux can be taken from
the measurements. Note that the simulated muon flux, the shape of their energy, and angular 
distributions are in a good agreement with Mei\&Hime prediction~\cite{mei} ($2.0\times10^{-7}$cm$^{-2}$s$^{-1}$)
 and the muon angular distribution agrees with
a measurement made by Ref.~\cite{measuredphi}.
FIG.~\ref{compPhi} shows the shape of muon azimuthal angle distribution compared with
a measurement~\cite{measuredphi} made by using an
active muon veto shield (room) at the Soudan Mine where the neutron detector situates inside.
The bin size of the simulated azimuthal angle is reorganized
according to that of the measurement data. A reasonable match is found which
demonstrates the reliability of the simulations. 
\par
\begin{figure}[htp!!!]
\includegraphics[width=0.48\textwidth]{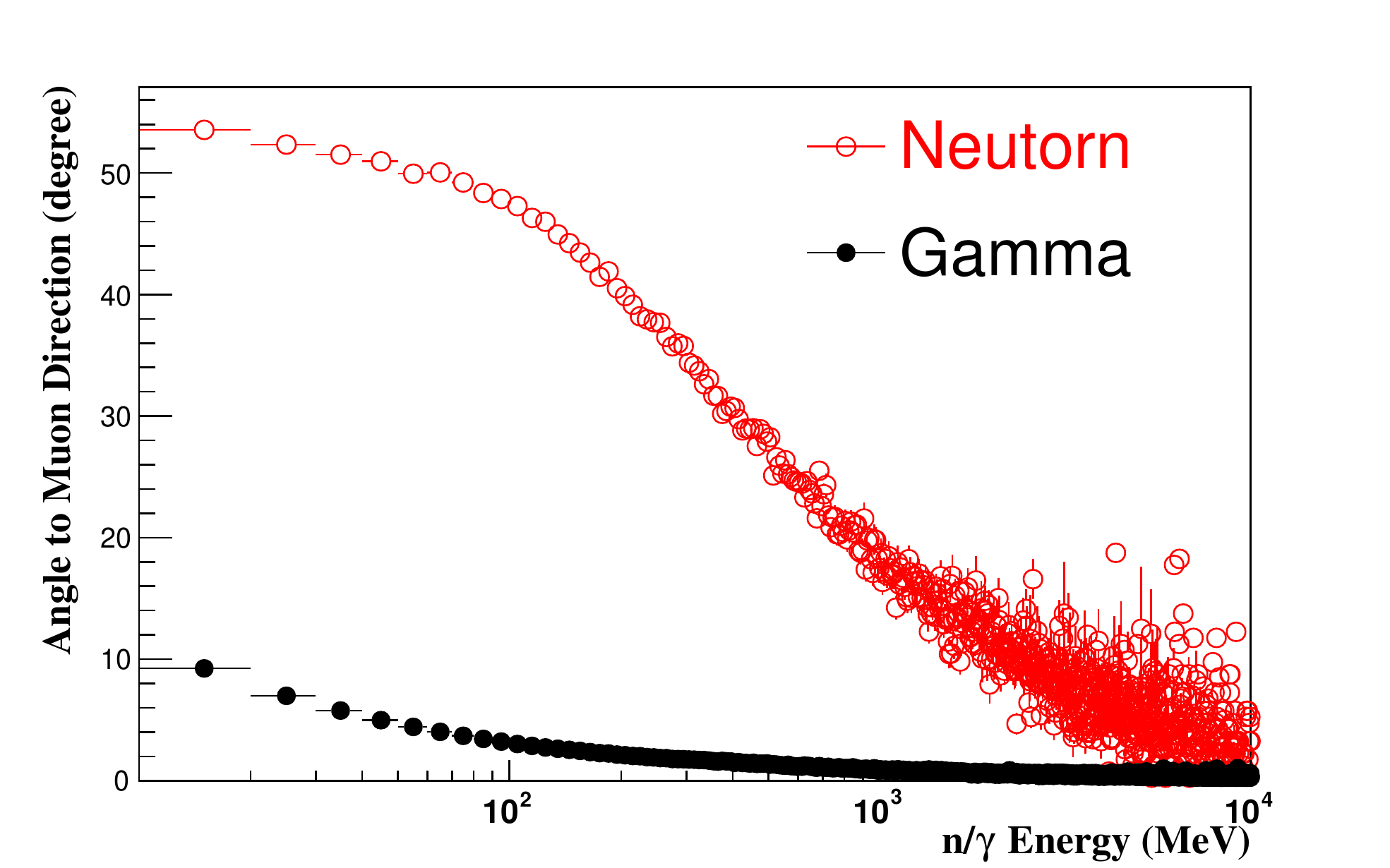}
\caption{\label{angleToMuon}
The angular correlation of secondary neurons/gammas to their primary muon.
The X-axis is the kinetic energy of secondary neutrons/gammas. The Y-axis represents
the average angle of $n/\gamma$ with respect to the direction of the primary muons.
}
\end{figure}

The angular correlation of shower particles with respect to the primary muon underground was studied
by many authors~\cite{mei, yfw, horn}. Their simulation results show that, in general, 
the neutrons with kinetic energy greater than 10 MeV are rather peak-forwarded along the muon track. 
In our simulation, muon shower information
is recorded on the ceiling and walls of a cavern with a size of 20 $m^{3}$ 
underground. In this case, the collected neutrons/gammas may not be produced
directly from their primary muon. Also, only a solid angle of $2\pi$ is considered,
which means no back scattering particles are recorded in the simulation. 
 The angular correlations of neutrons/gammas with respect
to parent muons are shown in FIG. \ref{angleToMuon}. As can be seen, the higher-energy secondaries including neutrons
are rather peak-forwarded  along the muon track and the lower-energy neutrons have a looser correlation.  

\subsection{Measurement Results}
\subsubsection{Muon flux}
\begin{figure*}[htp!!!]
\includegraphics[width=0.98\textwidth]{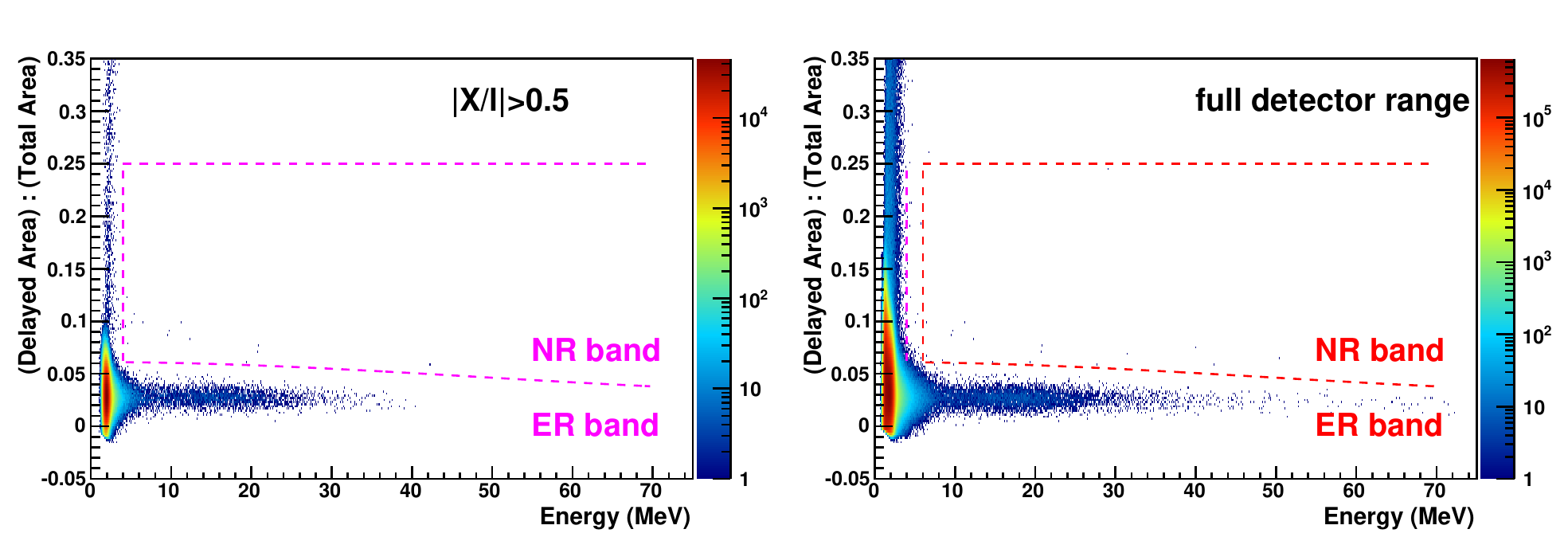}
\caption{\label{soudanResults}
Shown are the measurements of neutrons and gamma rays at Soudan Mine
 with a live time of 655.1 days.
The position of the events in the left plot are restricted to $|X/l|>0.5$.
The events in the right plot are from the entire detector.
The energy threshold for NR/ER separation is set to 4 MeV in the left plot
and 6 MeV in the right plot.
}
\end{figure*}

The detector has been taking data at the Soudan underground laboratory for
about two years. 
The detector response to nuclear recoil (NR) and electronic recoil (ER) is measured and shown
in FIG.~\ref{soudanResults}. For those nuclear recoils occurring
 far from the PMTs (i.e., the middle of the tube),
 the pulse shape difference (comparing with ER event) is easier to be washed out
due to the scattering and attenuation processes of photons on their way to the PMTs.
Therefore their energy threshold of NR/ER separation is
relatively higher than the events occurring  closer to the PMTs.
The left plot shows only the events with their position range $|X/l|>0.5$ in order to
get better NR/ER separation especially for low energies. A 4 MeV energy
threshold is set for good NR/ER separation in the left plot while
a 6 MeV energy threshold is applied in the right for
events from the entire detector.
The events from the nuclear recoil band (see FIG.~\ref{soudanResults})
are limited by the statistics. For those events with their visible energy below 4 MeV, the characterized
 ratio, $(Delayed Area) : (Total Area)$, suffers severely from the random noise which is
superimposed on the signal pulses.
\par
\begin{figure}[htp!!!]
\includegraphics[width=0.48\textwidth]{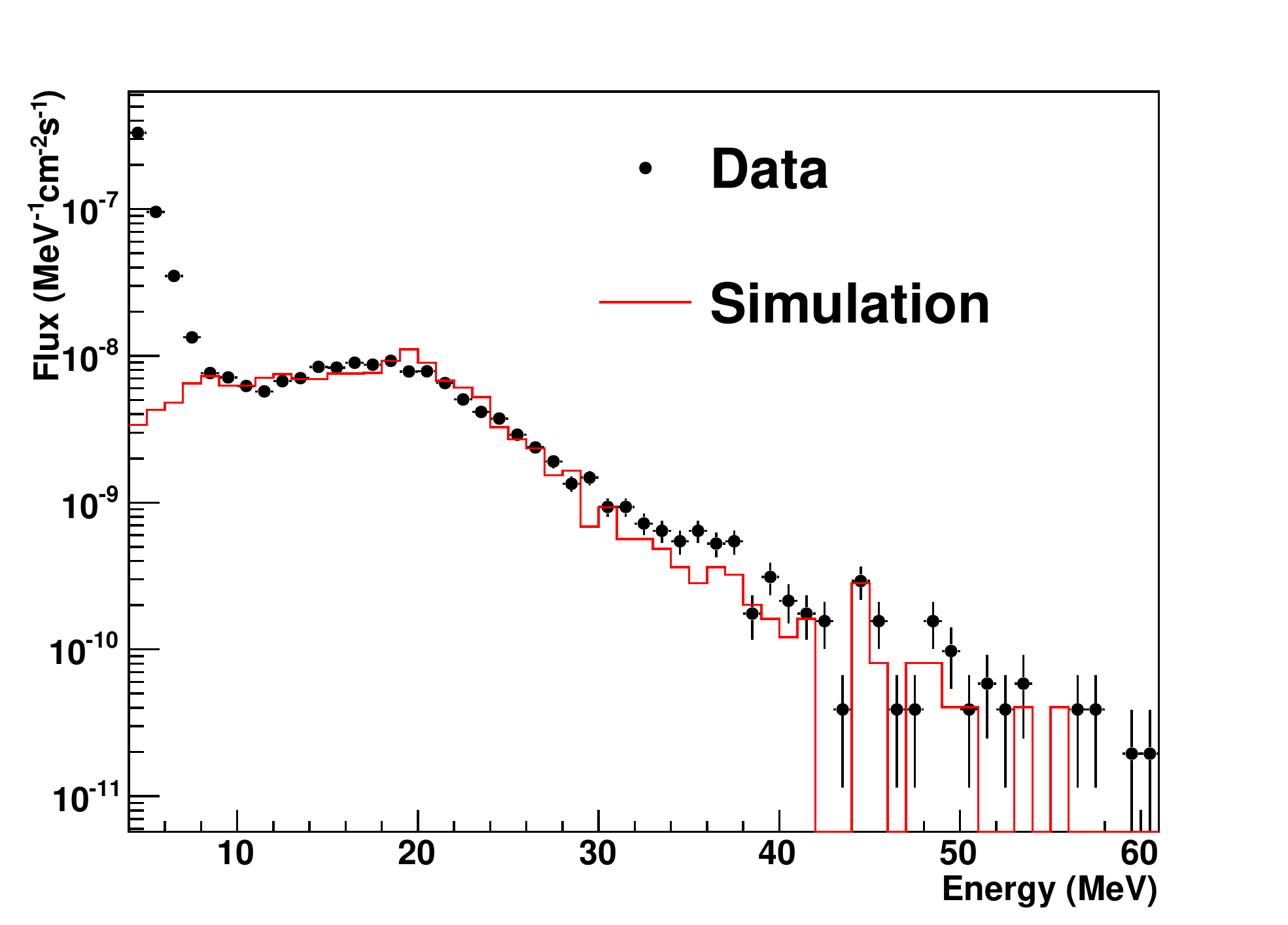}
\caption{\label{soudanMuon}
The energy spectrum of the energy depositions from electrons and muons (solid dots) obtained from the ER band 
in FIG.~\ref{soudanResults}. Detector response from the muon simulation (solid line)
is normalized, for the same live time, to the high energy region ($>$10 MeV) of the data. The normalization was performed
using the simulation divided by a factor of 1.21.
}
\end{figure}

The events from ER band in FIG.~\ref{soudanResults} (right plot) are picked out
and normalized according to 
their live time. The corresponding energy spectrum is displayed in FIG.~\ref{soudanMuon}
(solid dots).
Using the measured muons, the muon flux, passing through a horizontal surface,  is determined below.
\begin{equation}
\label{muonflux}
 \phi (E_{\mu}) = \frac{N_{\mu}}{t_{d}\cdot A_{d} \cdot \epsilon_{\mu}},
\end{equation}
where $N_{\mu}$, 13986 events with energy deposition greater than 10 MeV,
 is the number of muons across the entire detector, $t_{d}$ = 655.1 days, is the live time of the detector,
 $A_{d}$ = 1270 cm$^{2}$, represents the detector area (12.7 cm in diameter and 100 cm in length), 
and $\epsilon_{\mu}$ = 98\%,
stands for the detection efficiency of muons after taking into account the saturation of detector and the muon energy 
loss through radiative process. 
Thus, the muon flux is $\phi$ (E$_{\mu}$) = $1.99\times10^{-7}$cm$^{-2}$s$^{-1}$ which agrees with the Monte Carlo simulation.
However, $A_{d}$ = 1270 cm$^{2}$, a geometric cross-section of the detector, is not an effective area of the detector 
subtend to muons
with a pathlength greater than 5 cm in the detector for energy deposition greater than 10 MeV. This effective area must be
obtained through a Monte Carlo simulation.  
\par

Taking the muons and the associated secondaries in the cavern as the input of simulation, 
the detector response to muon showers in the cavern is obtained with the 
result normalized to ER data ($>10$ MeV), for the same live time, as shown in FIG.~\ref{soudanMuon} (red line). 
The simulated ER response to the muon shower is found to be a factor of 1.21 higher comparing to the 
measurement data. Taking this factor into account,  we found the effective area, $A_d$ = 1526 cm$^2$, which reduces
the muon flux from
 $1.99\times10^{-7}$cm$^{-2}$s$^{-1}$ to $1.65\times10^{-7}$cm$^{-2}$s$^{-1}$ while the 
shape of muon energy and angular distribution between the MC and the ER data agree very well as shown 
in FIG.~\ref{soudan_sim}.   
\subsubsection{Muon-induced neutron flux}
\begin{figure}[htp!!!]
\includegraphics[width=0.48\textwidth]{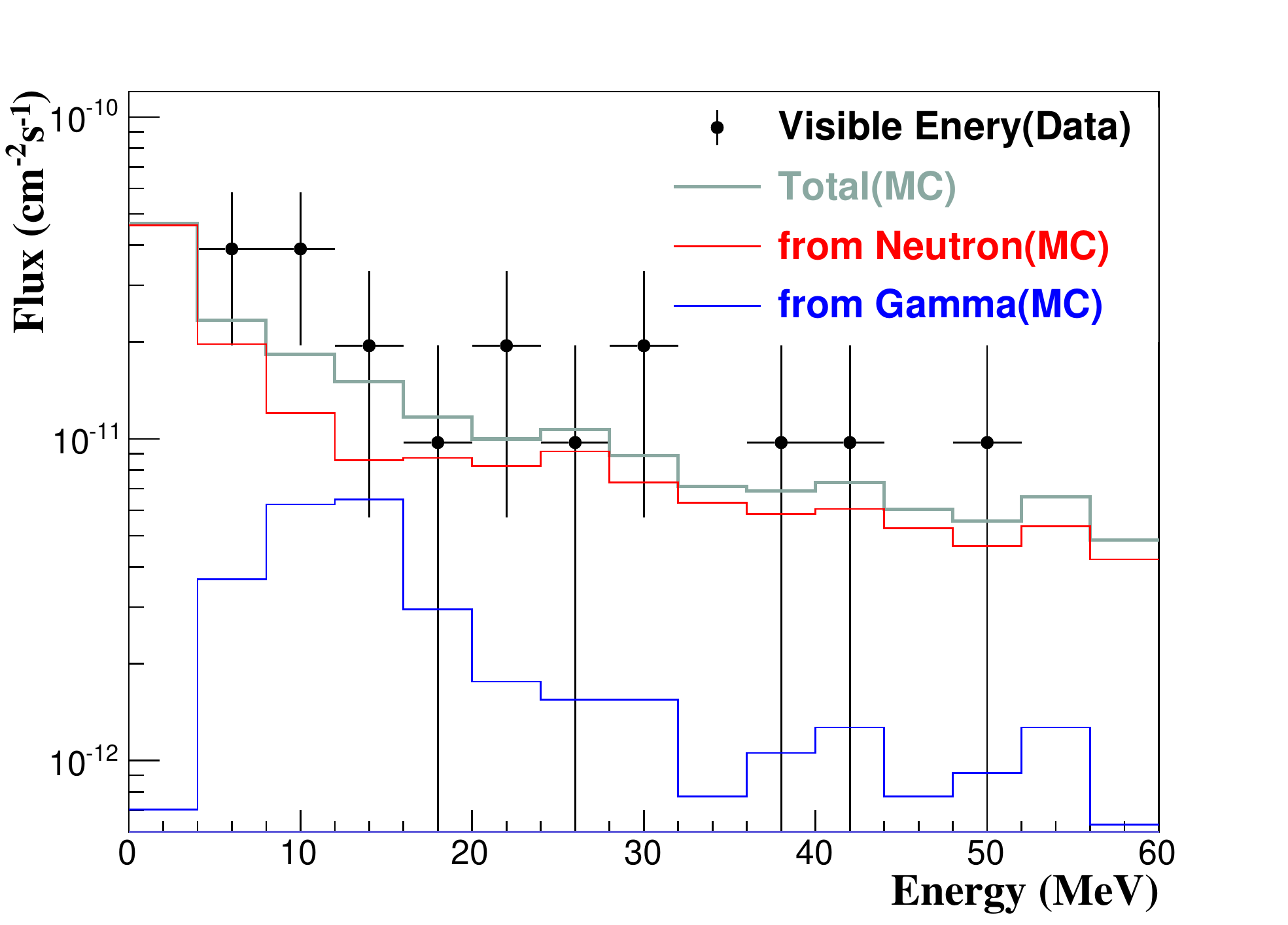}
\caption{\label{soudanNeutron}
The nuclear recoil events obtained from the NR band in FIG.~\ref{soudanResults}. 
The solid dots are the visible energies ``seen" by the PMTs. The solid blue and red lines 
stand for the nuclear recoils caused by gamma rays and neutrons, respectively.
The gray line is the simulated total NR response to the showers.   
A factor of $1/1.21$ is also applied to the simulated 
NR curves here in order to match the normalization assumed for the ER response. 
}
\end{figure}

Extracting the NR band from FIG.~\ref{soudanResults} (right plot), the
detector response to nuclear recoils is shown in FIG.~\ref{soudanNeutron}(solid dots).
Utilizing the selected neutrons from the NR band, the neutron flux can be obtained as follow.
\begin{equation}
\label{neuflux}
\phi (E_n)  = \int \frac{\frac{dN}{dE_n}}{t_d \cdot A_d \cdot \epsilon(E_n)} \cdot dE_n,
\end{equation}
where $\frac{dN}{dE_n}$ is the selected neutron events per energy bin, $\epsilon(E_n)$ is the detection efficiency 
which must be obtained through a Monte Carlo simulation since it depends  
strongly on neutron kinetic energy and scattering angle. 

The simulated detector response to nuclear recoils takes the input muon showers from
the results in FIG.~\ref{soudan_sim}.
The normalization factor of 1.21 is also applied to all simulated NR responses
(solid lines in FIG.~\ref{soudanNeutron}).
This reduces the calculated cosmogenic neutron flux, $\phi(E_n)$, from
$5.72\times10^{-9}$cm$^{-2}$s$^{-1}$ down to $4.73\times10^{-9}$cm$^{-2}$s$^{-1}$ in FIG.~\ref{soudan_sim}, for neutrons
with kinetic energy greater than 1 MeV.

Subsequently, the calculated cosmogenic gamma-ray flux is reduced from 8.57 $\times$10$^{-7}$cm$^{-2}s^{-1}$ to be
7.08 $\times$10$^{-7}$cm$^{-2}s^{-1}$.
The simulated NR contributions from neutrons (red line) and gamma rays (blue line) are separated.
It indicates that gamma rays induced nuclear recoils are comparable to that of neutrons in the
energy range of 10 to 15 MeV. In addition, the cosmogenic high-energy gamma rays are very penetrating and could undergo
photonuclear reaction to create neutrons inside the shield that is close to the detector. Therefore, the
cosmogenic high-energy gamma rays are an important
source of background to rare event physics experiments.
\par
\begin{figure}[htp!!!]
\includegraphics[width=0.48\textwidth]{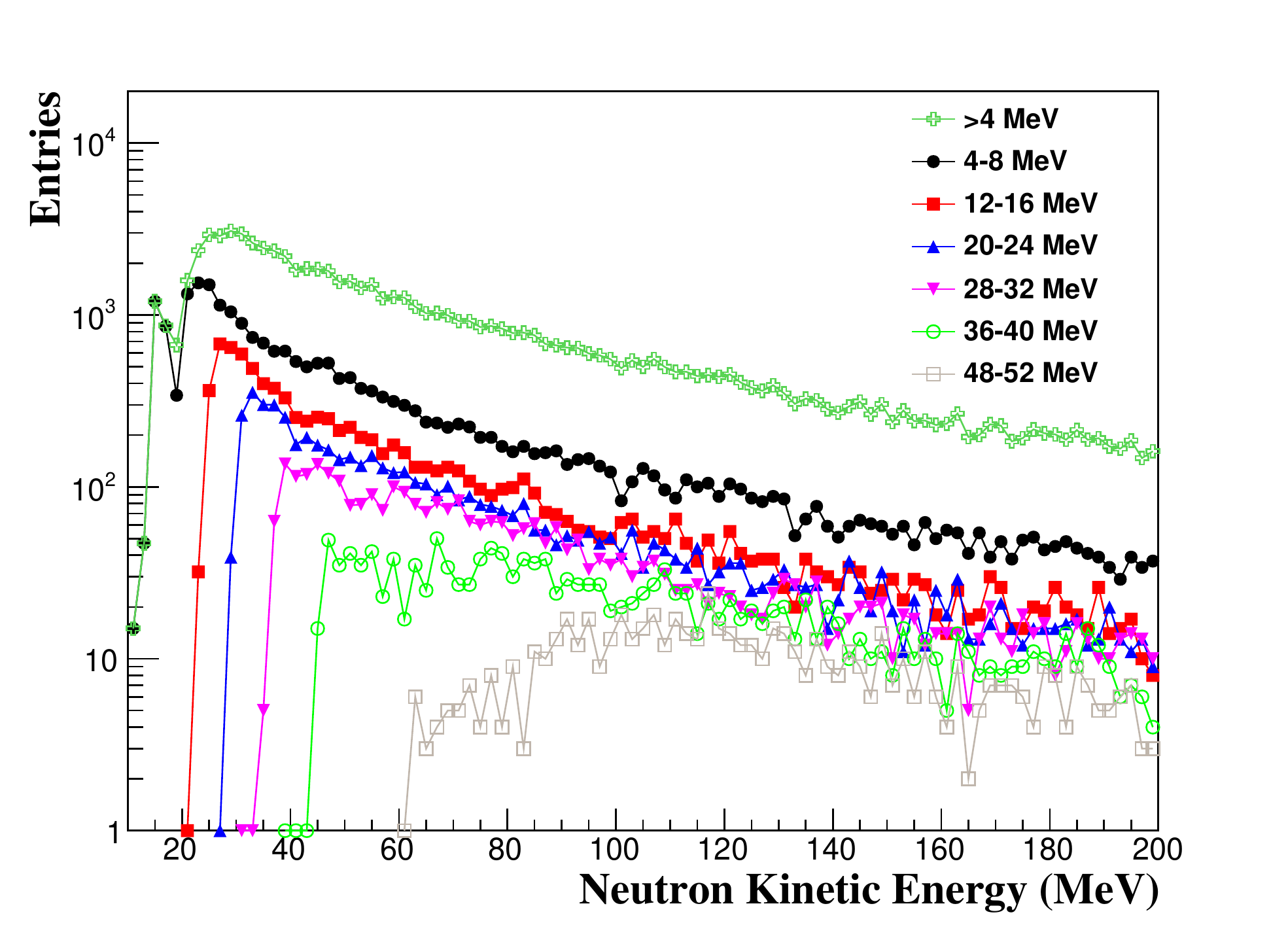}
\caption{\label{neuContribution}
The contribution of neutrons with different energies to the range of the measured energy depositions. 
}
\end{figure}

Total 24 nuclear recoil events have been observed (visible energy E$_n$$>$4 MeV) using the liquid scintillation
detector at Soudan Mine with the live time of 655.1 days. The visible recoil energies range from 4 MeV
to 50 MeV which corresponds to neutron energies from 20 MeV up to a few hundred MeV, depending on the scattering
angle. The contribution of neutrons with different energies to the range of the measured energy
depositions is studied through simulations (see FIG. \ref{neuContribution}). 
It shows that 4 to 8 MeV energy depositions are mainly contributed by $\sim$20 MeV neutrons. 
The little peak around 15 MeV is caused by $C(n, 3\alpha)$ processes in the detector.  
The results indicate 
that the energy correlation between recoils and incident neutrons is obvious for low energies. 
The correlation becomes more obscure as energy rising.
\par
\begin{figure}[htp!!!]
\includegraphics[width=0.48\textwidth]{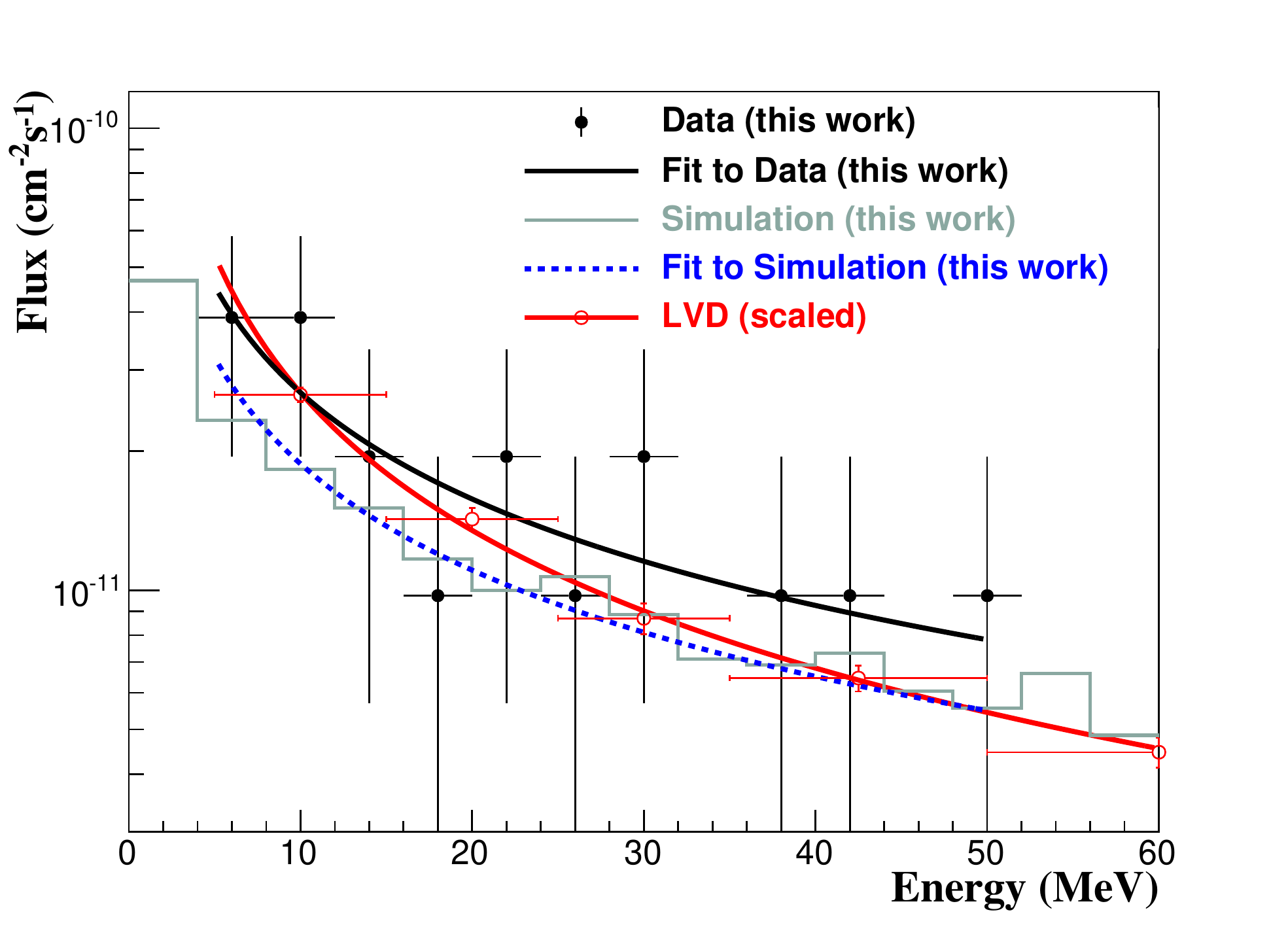}
\caption{\label{compLVD}
The measured nuclear recoil spectrum at Soudan Mine (black solid dots) is compared with the one
measured by LVD experiment at Gran Sasso (red open dots). The total flux of the LVD measurement is 
normalized by a factor of $1.66\times10^{-5}$ in the plot. The gray line is the simulation
result for the Soudan measurement. For fitting results, black line for Soudan data, dashed line 
for the simulation and 
red line for LVD measurement, are also presented, respectively. 
}
\end{figure}

Muon induced fast neutrons underground have been reported by LVD experiment\cite{lvd} at 
Gran Sasso depth (3.1 km.w.e).
The observed proton recoil spectra is up to 300 MeV, corresponding to neutron energy up to a GeV 
depending on the scattering angle,  with much higher
statistics due to a much bigger detector. The measured recoil spectrum is parameterized by a function
$dN/dE = A\cdot E^{-\alpha}$ with $A = (1.58\pm0.14)\cdot 10^{-5}$ neutrons($\mu$$^{-1}$counter$^{-1}$MeV$^{-1}$ and 
$\alpha = 0.99\pm0.02$. The spectrum and the fitted curve are normalized by a factor of $1.66\times10^{-5}$
and presented in FIG. \ref{compLVD} in order to do a shape comparison with our results. 
A similar fitting
is also applied to our data with $A = (1.56\pm1.65)\cdot 10^{-10}$ and $\alpha = 0.77\pm0.38$.   
The distribution of the Soudan data from this work is more flat than that of Gran Sasso from LVD. 
This difference is attributed to the
difference of the depth between two sites.  
Note that the simulated nuclear recoil curve in FIG.~\ref{neuContribution} and~\ref{compLVD} 
takes a normalization factor of 1.21 from the muon measurement (see FIG.~\ref{soudanMuon}). 
Fitted with the same function to the simulation curve gives the parameter 
$A = 1.10\cdot 10^{-10}$ and $\alpha = 0.76$. As can be seen in Fig.~\ref{compLVD}, 
the difference between the fitted curve with the measured 
neutron data and the fitted curve from the adjusted simulation results, using muon data, 
 is about 30\%. 
\par
We integrate the energy region above the 4 MeV energy threshold, the measured neutron flux is determined to be
2.23$\times$10$^{-9}$cm$^{-2}$s$^{-1}$. Considering the 4 MeV energy threshold accounts for 
neutrons with energy above 20 MeV, the measured neutron flux, 2.23$\times$10$^{-9}$cm$^{-2}$s$^{-1}$, corresponds
to neutrons with energy greater than 20 MeV. Similarly, the simulated neutron flux, for neutron energy greater than 20 MeV,
is determined to be 1.90$\times$10$^{-9}$cm$^{-2}$s$^{-1}$.   
\par
\begin{figure}[htp!!!]
\includegraphics[width=0.48\textwidth]{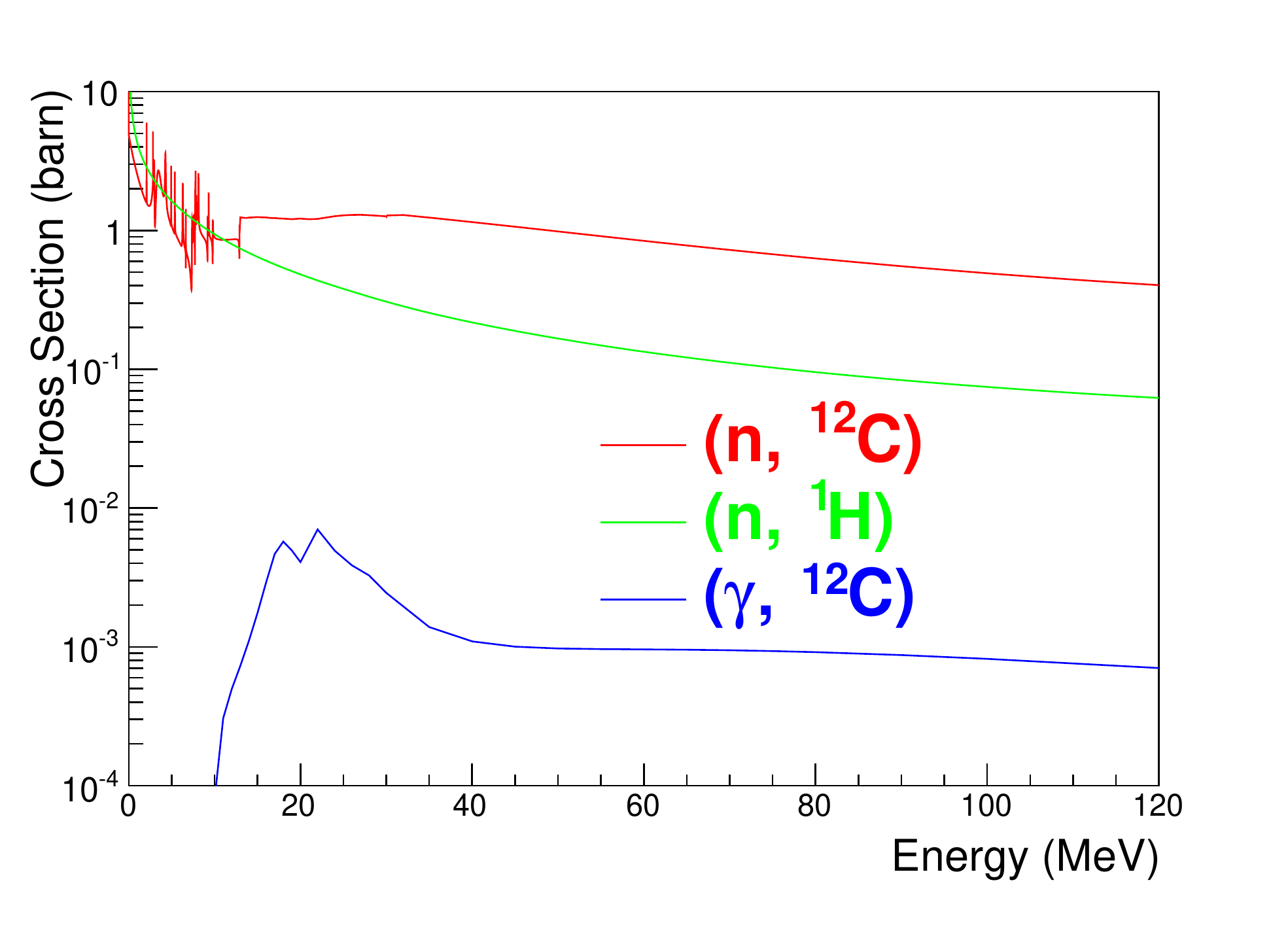}
\caption{\label{xsComp}
The comparison of the interaction cross sections~\cite{endf}
for neutron and gamma ray in liquid scintillators.
}
\end{figure}

Muon electromagnetic showers produce a large amount of secondary particles which
can contribute to nuclear recoils in liquid scintillators. 
General speaking, neutron elastic/inelastic processes are considered to be 
the dominated NR interaction
channels comparing to muon-nucleus and photo-nuclear interactions. 
The simulation finds that muon showers contain a factor of 100 more gamma rays than neutrons.  
Considering this higher intensity and the photonuclear interaction cross-section in liquid scintillators, 
we evaluate the contribution to nuclear recoil, from these secondary high-energy gamma rays, in the detector.
As shown in FIG.~\ref{xsComp}, gamma rays break $^{12}$C nucleus 
through reaction of $^{12}$C($\gamma, 3\alpha)$ when the energy of gamma ray exceeding 13 MeV. The contribution 
from $^{12}$C is less than 0.1\% compared to protons above 10 MeV, while the contribution from those $\alpha$ recoils are 
in the level of less than 1.0\% in the NR band, according to the Monte Carlo simulation.
The NR events produced through muon-nucleus processes
are found to be negligible in the data.     

\section{Discussion and Conclusion}
The muons at the Earth's surface are utilized to calibrate the detector response to
high energies. The result is a reasonable approximation by assuming a linear
relation between the light output and the energy deposition at high energies.  
Neutrons at the Earth's surface are measured using
the 12-liter liquid scintillation detector. The discrepancy between the measured data
and the simulated result above 20 MeV is within a factor of 2. A full calibration
using high energy neutron beam would help in terms of the energy and position 
calibrations for high energies so that we can confidently identify 
the sources of issue, either
from the detector response itself or from the external input.    
\par
Muon and neutron fluxes at Soudan Mine are simulated by taking into account the surface 
mountain profile and rock density. Since the reality of the rock configuration is 
more complicated, the simulated results have to be adjusted by the measurement in 
terms of their absolute fluxes.
The lower end of measured visible energy from neutron band is limited by the random noise
($\sim$4 MeV). Given the quenching effect and the energy transfer from neutron-ion
collisions, the measured neutrons should have kinetic energies above $\sim$20 MeV
 according to this 4 MeV threshold. The highest
visible energy of $\sim$50 MeV in the neutron band accounts for the 
potential incident neutrons with energy about a few hundred MeV depending on the scattering angle.   
\par
The statistical error of the muon measurement is about 1.23\% while it is 22.94\% on the average for the neutron flux
due to the relatively lower intensity at the mine. The dominant systematic uncertainty for muons comes
from a 5\% variation in total muon rates observed in the experiment. In addition, the energy and position calibration
contributed to about 3\% uncertainty determined using the Monte Carlo simulation. 
For neutrons, in addition to the inherited systematics from the parent muons, the systematic uncertainty is 
mainly from the normalization factor which contributes 30\%. The uncertainty, at level of about 6\% to the total flux, 
induced by the quenching factor matrix, 
 is determined by the Monte Carlo simulation.
Adding the sources of systematic uncertainties in quadrature,  
the final muon flux (E$_{\mu}$ $>$ 1 GeV) is determined to be 
($1.65\pm 0.02 (sta.) \pm 0.1 (sys.) ) \times10^{-7}$ cm$^{-2}s^{-1}$ and 
the measured neutron flux (E$_{n}$ $>$ 1 MeV) is 
($2.23 \pm 0.52 (sta.) \pm 0.99 (sys.) ) \times10^{-9}$ cm$^{-2}s^{-1}$.    
The former is in a good agreement with the previous measurement made by the MINOS far detector~\cite{minos1}. 
The latter agrees with the prediction made by Ref.~\cite{mei} reasonably well. Table~\ref{comparison} summarizes 
a comparison between Ref.~\cite{mei} and this work. 
\begin{table}
\caption{A comparison between Ref.~\cite{mei} and this work.}
\begin{tabular}{l|l|l|r}
\hline
 Sources & Muon flux  & \multicolumn{2}{c}{Neutron flux}  \\ \hline
         &(cm$^{-2}$s$^{-1}$) & \multicolumn{2}{c}{(cm$^{-2}$s$^{-1}$)} \\ \hline
         &E$_{\mu}$$>$ 1 GeV & E$_{n}$$>$ 1 MeV & E$_{n}$$>$20 MeV \\
\hline
Mei\&Hime~\cite{mei}	&2.0$\times$ 10$^{-7}$ & 5.84$\times$10$^{-9}$ & 2.5$\times$10$^{-9}$ \\
MC (this work)	& 1.99$\times$10$^{-7}$	& 5.72$\times$10$^{-9}$ &1.9$\times$10$^{-9}$ \\
Data (this work) &1.65$\times$10$^{-7}$ &  &2.23$\times$10$^{-9}$ \\
\hline
\end{tabular}
\label{comparison}
\end{table}
\par
As can be seen in Table~\ref{comparison}, the Monte Carlo simulation with FLUKA~\cite{fluka} implemented by Mei\&Hime~\cite{mei}
agrees reasonably well with the Monte Carlo simulation with GEANT4
 (GEANT4.9.5.p02 + Shielding module physics list) performed by this work. The difference between 
two simulations and the measurements made by this work is about 30\%, which is mainly due to the variation of rock density in reality.
Thus, our measurements provides a benchmark for both FLUKA and GEANT4 simulations. 
\par
In summary, we have demonstrated the capability of detecting cosmogenic 
neutrons with a 12-liter liquid scintillation detector. 
Although there are only 24 events (E$_n$$>4$ MeV) detected in 
the NR band
in two years,
they are well separated from electron recoils. Due to relatively low cost of the whole detector, 
an array of hundreds of such detectors would be able to collect the sufficient
 statistics for studying cosmogenic neutrons in the underground laboratory to benchmark Monte Carlo simulation tools.    
\section{Acknowledgement}
The authors wish to thank Fred Gray, 
Keenan Thomas,  
Anthony Villano and Priscilla Cushman for their invaluable suggestions and help. 
We would also like to thank Christina Keller, Angela Chiller, and Wenzhao Wei for careful reading of this manuscript.
This work was supported in part by NSF PHY-0758120, PHYS-0919278, PHYS-0758120,
PHYS-1242640, DOE grant DE-FG02-10ER46709, 
the Office of Research at the University
of South Dakota and a 2010 research center support by the State of South Dakota.
Computations supporting this project were performed on High Performance Computing systems at 
the University of South Dakota.


\begin{thebibliography}{99}
\bibitem{trimble} V. Trimble, Annual Review of Astronomy and Astrophysics, \textbf{25}, 425 (1987). 
\bibitem{dama} R. Bernabei et al. (DAMA Collaboration), Riv. N. Cim., \textbf{26}, 1 (2003).
\bibitem{cdms} Z. Ahmed et al. (CDMS II Collaboration), Science, \textbf{327}, 1619 (2010).
\bibitem{cogent} C. E. Aalseth et al. (CoGeNT Collaboration), Phys. Rev. Lett., \textbf{107}, 141301, (2011).
\bibitem{cresst} G. Angloher et al. (CRESST-II Collaboration), Eur. Phys. J. C \textbf{72}, 1971 (2012).
\bibitem{cdms2} Z. Ahmed et al. (CDMS and EDELWEISS Collaborations), Phys. Rev. D \textbf{84}, 011102 (2011). 
\bibitem{xenon100} E. Aprile et al. (XENON Collaboration), Phys. Rev. Lett. \textbf{109}, 181301 (2012). 
\bibitem{lux} D. S. Akerib et al. (LUX collaboration), Phys. Rev. Lett. \textbf{112}, 091303 (2014); arXiv:1310.8214.
\bibitem{supercdms} R. Agnese et al. (SuperCDMS Collaboration), Phys. Rev. Lett. 112 (2014) 241302, arXiv:1402.7137.
\bibitem{picasso} M. Barnabe-Heider et al. (PICASSO), Phys. Lett. B \textbf{624}, 186 (2005). 
\bibitem{naiad} G. J. Alner et al. (UKDMC), Phys. Lett. B \textbf{616}, 17 (2005).
\bibitem{zeplin} V. A. Kudryavtsev (UKDMC), in the Fifth International Workshop on the Identification
		of Dark Matter, Edinburgh, Scotland, 2004.
\bibitem{edelweiss} A. Benoit et al. (EDELWEISS Collaboration), Phys. Lett. B \textbf{616}, 25 (2005). 
\bibitem{simple} T. A. Girard et al., Phys. Lett. B \textbf{621}, 233 (2005). 
\bibitem{superKamiokande} S. Desai et al. (Super-Kamiokande Collaboration), Phys. Rev. D \textbf{70}, 083523 (2004).
\bibitem{mei} D.-M. Mei and A. Hime, Phys. Rev. D \textbf{73}, 053004 (2006).
\bibitem{yfw} Y.-F. Wang, et al., Physical Review D \textbf{64}, 013012 (2001).
\bibitem{vku} V. Kudryavtsev, N. Spooner, J. McMillan, Nuclear Instruments and Methods in Physics Research A 
	\textbf{505}, 688 (2003).
\bibitem{hav} H. Araujo, V. Kudryavtsev, N. Spooner, T. Sumner, Nuclear Instruments and Methods in 
	Physics Research A \textbf{545}, 398 (2005).
\bibitem{vkl} V. Kudryavtsev, L. Pandola, V. Tomasello, The European Physical Journal A \textbf{36}, 171 (2008).
\bibitem{ggv} G. Gorshkov, V. Zyabkin, R. Yakovlev,  Soviet Journal of Nuclear Physics, \textbf{18}, 57 (1974).
\bibitem{mma} M. Marino, et al.,  Nuclear Instruments and Methods in Physics Research A \textbf{582}, 611 (2007).
\bibitem{har} H. Araujo, et al., Astroparticle Physics \textbf{29}, 471 (2008).
\bibitem{sbe} S. Abe, et al.,  Physical Review C \textbf{81}, 025807 (2010).
\bibitem{borx} G. Bellini et al.,  JINST, \textbf{6}, 05005 (2011).
\bibitem{lre} L. Reichhart et. al. Astropart. Physics \textbf{47}, 67 (2013).
\bibitem{edelweiss2} B. Schmidt et. al., Astropart. Physics \textbf{44}, 28 (2013).
\bibitem{lvd} M. Aglietta et al, Proc. of the 26th ICRC, Salt Lake City, Vol 2, 1999; arXiv:hep-ex/9905047.
\bibitem{cecil} R. A. Cecil B. D. Anderson and R. Madey, Nucl. Instr. and
		Meth., \textbf{161}, 439 (1979).
\bibitem{batch} R. Batchelor, W. B. Gilboy, J. B. Parker and J. H. Towle, 
		Nucl. Instr. and Meth., \textbf{13}, 70 (1961).
\bibitem{verb} V. V. Verbinski, W. R. Burrus, T. A. Love, W. Zobel and N. W. Hill,
		Nucl. Instr. and Meth. \textbf{65}, 8 (1968).
\bibitem{gul} K. Gul, A. A. Naqvi and H. A. Al Juwair, Nucl. Instr. and
		Meth. A \textbf{278}, 470 (1989).
\bibitem{aksoy} A. Aksoy et. al., Nucl. Instr. and Meth. A \textbf{337}, 486 (1994).
\bibitem{nakao} N. Nakao et. al., Nucl. Instr. and Meth. A \textbf{362}, 454 (1995).
\bibitem{o5s} R. E. Textor and V. V. Verbinski, ORNL-4160, Oak Ridge 
		National Laboratory (1968).
\bibitem{scinful} J. K. Dickens, ORNL-6463, Oak Ridge National Laboratory (1988).
\bibitem{LSneu} C. Zhang, et. al., Nucl. Instr. and Meth. A \textbf{729}, 138 (2013).
\bibitem{birkslaw} J. B. Birks, The Theory and Practice of Scintillation Counting, (Pergamon, New York, 1964). 
\bibitem{geant4} S. Agostinelli, et. al., Nucl. Instr. and Meth. A \textbf{506}, 250 (2003); 
		K. Amako et al., IEEE Transactions on Nuclear Science \textbf{53}, 270 (2006).
\bibitem{modGaisser} M. Guan, et. al., http://escholarship.org/uc/item/6jm8g76d.
\bibitem{elevation} http://www.city-data.com/city/Vermillion-South-Dakota.html.
\bibitem{neuNYdata} M. S. Gordon, et. al., IEEE Transactions on Nuclear Science \textbf{51}, 3427 (2004). 
\bibitem{birks} J. B. Birks and F. A. Black, Proc. Phys. Soc. A \textbf{64}, 511i (1951).
\bibitem{quenchingfactor} G. Bruno, JINST \textbf{8}, T05004 (2013).
\bibitem{ej301qf} H. Wan Chan Tseung, J. Kaspar, N. Tolich, Nucl. Instr. and Meth. A \textbf{654}, 318 (2011).
\bibitem{estar} http://physics.nist.gov/PhysRefData/Star/Text/ESTAR .html
\bibitem{pstar} http://physics.nist.gov/PhysRefData/Star/Text/PSTAR .html
\bibitem{astar} http://physics.nist.gov/PhysRefData/Star/Text/ASTAR .html
\bibitem{minos} P. Adamson et al. (MINOS Collaboration), Phys. Rev. D \textbf{73}, 072002 (2006). 
\bibitem{cdms0} D. S. Akerib et al. (CDMS Collaboration), Phys. Rev. Lett. \textbf{93}, 211301 (2004).
\bibitem{cogent0} C. E. Aalseth et al. (CoGeNT Collaboration), Phys. Rev. Lett. \textbf{106}, 131301 (2011).
\bibitem{minos1} P. Adamson et al. (MINOS Collaboration), Phys. Rev. D \textbf{81}, 012001 (2010).
\bibitem{usgs} http://srtm.csi.cgiar.org/.
\bibitem{location} http://homepages.spa.umn.edu/$\sim$schubert/far/s2rock/ position.html
\bibitem{soudanRock} K. Ruddick, Underground Particle Fluxes in the Soudan Mine, Internal Note NuMI-L-210, 1996.
\bibitem{measuredphi} N. Pastika, Bachelors Thesis, University of Minnesota, 2008.
\bibitem{horn} M. Horn, Ph.D. thesis, Universit\"{a}t Karlsruhe, 2007.
\bibitem{musun} V. A. Kudryavtsev, Computer Physics Communications, \textbf{180}, 339 (2009).
\bibitem{deuterons} M. Moszy\'{n}ski, et al., Nucl. Instr. and Meth. A \textbf{343}, 536 (1994). 
\bibitem{endf} M.B. Chadwick et al., Nucl. Data Sheets \textbf{112}, 2887 (2011).
\bibitem{fluka} A. Fass´o, A. Ferrari, P.R. Sala, “Electron-photon transport in FLUKA: status”, Proc. of the Monte Carlo 2000
Conference (Lisbon, October 23-26, 2000), A. Kling, F. Barao, M. Nakagawa, L. Tavora, P. Vaz - eds. (Springerverlag, Berlin, 2001),
 p.159-164; A. Fass´o, A. Ferrari, J. Ranft, P.R. Sala, “FLUKA: Status and Prospective for Hadronic Applications”, 
Proceedings of the Monte Carlo 2000 Conference (Lisbon, October 23-26, 2000), A.Kling, F.Barao, M.Nakagawa, 
L.Tavora, P.Vaz - eds., Springer-Verlag Berlin, p.955-960 (2001). 
\end{thebibliography}
\end{document}